\documentclass[longauth]{aa}
\usepackage{color}
\usepackage{url}
\usepackage{natbib}
\usepackage[switch]{lineno}
\pdfmapfile{+txfonts.map}
\usepackage{caption,subcaption}
\usepackage{graphicx}
\usepackage{graphics}
\usepackage{amssymb}
\usepackage{amsmath}
\usepackage{float}
\usepackage{txfonts}
\begin{document}

   \title{Multi-wavelength characterization of the blazar S5~0716+714  during an unprecedented outburst phase}

   \subtitle{ }
% authors 23.11.2017  Format AA
%
\author{
MAGIC Collaboration: 
M.~L.~Ahnen\inst{1} \and
S.~Ansoldi\inst{2,20} \and
L.~A.~Antonelli\inst{3} \and
C.~Arcaro\inst{4,27} \and
D.~Baack\inst{5} \and
A.~Babi\'c\inst{6} \and
B.~Banerjee\inst{7} \and
P.~Bangale\inst{8} \and
U.~Barres de Almeida\inst{8,9} \and
J.~A.~Barrio\inst{10}
% ...
}
\institute { ETH Zurich, CH-8093 Zurich, Switzerland
\and Universit\`a di Udine, and INFN Trieste, I-33100 Udine, Italy
\and National Institute for Astrophysics (INAF), I-00136 Rome, Italy
\and Universit\`a di Padova 
\and Technische Universit\"at Dortmund, D-44221 Dortmund, Germany
\and Croatian MAGIC Consortium: University of Rijeka, 51000 Rijeka, University of Split - FESB, 21000 Split,  University of Zagreb - FER, 10000 Zagreb, University of Osijek, 31000 Osijek and Rudjer Boskovic Institute, 10000 Zagreb, Croatia.
\and Saha Institute of Nuclear Physics, HBNI, 1/AF Bidhannagar, Salt Lake, Sector-1, Kolkata 700064, India
\and Max-Planck-Institut f\"ur Physik, D-80805 M\"unchen, Germany
\and now at Centro Brasileiro de Pesquisas F\'isicas (CBPF), 22290-180 URCA, Rio de Janeiro (RJ), Brasil
\and Unidad de Part\'iculas y Cosmolog\'ia (UPARCOS), Universidad Complutense, E-28040 Madrid, Spain
\and Inst. de Astrof\'isica de Canarias, E-38200 La Laguna, and Universidad de La Laguna, Dpto. Astrof\'isica, E-38206 La Laguna, Tenerife, Spain
\and University of \L\'od\'z, Department of Astrophysics, PL-90236 \L\'od\'z, Poland
\and Deutsches Elektronen-Synchrotron (DESY), D-15738 Zeuthen, Germany
\and Institut de F\'isica d'Altes Energies (IFAE), The Barcelona Institute of Science and Technology (BIST), E-08193 Bellaterra (Barcelona), Spain
\and Universit\`a  di Siena and INFN Pisa, I-53100 Siena, Italy
\and Universit\"at W\"urzburg, D-97074 W\"urzburg, Germany
\and Finnish MAGIC Consortium: Tuorla Observatory and Finnish Centre of Astronomy with ESO (FINCA), University of Turku, Vaisalantie 20, FI-21500 Piikki\"o, Astronomy Division, University of Oulu, FIN-90014 University of Oulu, Finland
\and Departament de F\'isica, and CERES-IEEC, Universitat Aut\'onoma de Barcelona, E-08193 Bellaterra, Spain
\and Universitat de Barcelona, ICC, IEEC-UB, E-08028 Barcelona, Spain
\and Japanese MAGIC Consortium: ICRR, The University of Tokyo, 277-8582 Chiba, Japan; Department of Physics, Kyoto University, 606-8502 Kyoto, Japan; Tokai University, 259-1292 Kanagawa, Japan; The University of Tokushima, 770-8502 Tokushima, Japan
\and Inst. for Nucl. Research and Nucl. Energy, Bulgarian Academy of Sciences, BG-1784 Sofia, Bulgaria
\and Universit\`a di Pisa, and INFN Pisa, I-56126 Pisa, Italy
\and Humboldt University of Berlin, Institut f\"ur Physik D-12489 Berlin Germany
\and also at Dipartimento di Fisica, Universit\`a di Trieste, I-34127 Trieste, Italy
\and also at Port d'Informaci\'o Cient\'ifica (PIC) E-08193 Bellaterra (Barcelona) Spain
\and also at INAF-Trieste and Dept. of Physics \& Astronomy, University of Bologna
\and INFN, I-35131 Padova, Italy
% ...
}

   \date{}

% \abstract{}{}{}{}{} 
% 5 {} token are mandatory
%Context
  \abstract{The BL~Lac object S5~0716+714, a highly variable blazar, underwent an impressive outburst in January 2015 (Phase A), followed by minor activity in February (Phase B). The MAGIC observations were triggered by the optical flux observed in Phase A, corresponding to the brightest ever reported state of the source in the R-band. 
  }
%Aims
  {The comprehensive dataset collected is investigated in order to shed light on the mechanism of the broadband emission.}
%Methods  
{Multi-wavelength light curves  have been studied together with the broadband Spectral Energy Distributions (SEDs). The data set collected spans from radio (Effelsberg, OVRO, Mets\"{a}hovi, VLBI, CARMA, IRAM, SMA), UV (\textit{Swift}-UVOT), optical photometry and polarimetry (Tuorla, Steward, RINGO3, KANATA, AZT-8+ST7, Perkins, LX-200), X-ray (\textit{Swift}-XRT and \textit{NuSTAR}), high-energy (HE, 0.1 GeV < E < 100 GeV) with \textit{Fermi}-LAT to the very-high-energy (VHE, E>100 GeV) with MAGIC.}
%Results
{The flaring state of Phase A was detected in all the energy bands, providing for the first time a multi-wavelength sample of simultaneous data from the radio band to the VHE. In the constructed SED the \textit{Swift}-XRT+\textit{NuSTAR} data constrain the transition between the synchrotron and inverse Compton components very accurately, while the second peak is constrained from 0.1~GeV to 600~GeV by \textit{Fermi}+MAGIC data. The broadband SED cannot be described with a one-zone synchrotron self-Compton model as it severely underestimates the optical flux in order to reproduce the X-ray to $\gamma$-ray data. Instead we use a two-zone model. The electric vector position angle (EVPA) shows an unprecedented fast rotation. An estimation of the redshift of the source by combined HE and VHE data provides a value of $z = 0.31 \pm 0.02_{stats} \pm 0.05_{sys}$, confirming the literature value.} 
%Conclusions
{The data show the VHE emission originating in the entrance and exit of a superluminal knot in and out a recollimation shock in the inner jet. A shock-shock interaction in the jet seems responsible for the observed flares and EVPA swing. This scenario is also consistent with the SED modelling.}

   
  \titlerunning{MWL characterization of S5~0716+714 during an unprecedented outburst}
  \authorrunning{M.~L.~Ahnen et~al.} 

   \maketitle
\section{Introduction}
%\linenumbers

The blazar S5~0716+714 is a BL Lac object characterized by extreme variability in almost all energy bands. Because of the featureless optical continuum \citep{paiano2017} it is hard to estimate its redshift. \cite{nilsson2008} claimed a value of $z = 0.31 \pm 0.08$ based on the photometric detection of the host galaxy. Detection of intervening Ly$\alpha$ systems in the ultra-violet spectrum of the source confirms the earlier estimates with a redshift value $z < 0.32$ (95\% confidence) \citep{danforth2013}.\par
The familiar shape of the Spectral Energy Distribution (SED) of blazars, in which the two bumps are identified as synchrotron and inverse Compton (IC) respectively, is used for their classification. In the case of S5~0716+714 the first peak of the SED is located between $10^{14}$ and $10^{15}$ Hz, leading to a classification as intermediate synchrotron peaked blazar or IBL (Intermediate-peaked BL Lac object) \citep{giommi1999,ackermann2011}.\par
Because of its remarkable variability, S5~0716+714 has been the subject of many optical monitoring campaigns  \citep{wagner1996,  montagni2006, rani2011, rani2013a, rani2015}. Several authors carried out flux variability studies \citep[e.g.][]{quirrenbach1991,wagner1996} and morphological/kinematic studies at radio frequencies \citep{antonucci1986, witzel1988, jorstad2001, bach2005, rastorgueva2011}. The observed intraday variability at radio wavelengths is likely to be a mixture of intrinsic and external (due to interstellar scintillation) mechanisms. \cite{wagner1996} reported a significant correlation between optical -- radio flux variations at day-to-day timescales.  \cite{rani2010b} reported the detection of a $\sim~15$-min quasi-periodic oscillations (QPOs) at optical frequencies, so far the shortest ones ever observed in any blazar.\par
The study of the broadband flux variability is a difficult task because of the complexity of the flaring activity, which can vary rapidly on a timescale of a few hours to days, while on a timescale of $\sim~1$ year a broader and slower variability trend has been reported \citep{raiteri2003,rani2013a}. As seen by Very Long Baseline Interferometry (VLBI) studies \citep{bach2005, britzen2009}, the source presents a core-dominated jet pointing towards the north. Very Large Array observations show a halo-like jet misaligned with it by $\sim$90$^{\circ}$ on kiloparsec scales \citep{antonucci1986}: \cite{britzen2009} suggest that there is an apparent stationarity of jet components relative to the core. However, recently apparent speeds of $\sim~$40~c have been reported \citep{rastorgueva2011, larionov2013, lister2013, rani2015}. \citep{britzen2009, rastorgueva2011, rani2014} observed non-radial motion and wiggling component trajectories in the inner milliarcsecond jet region.\par 
Observations by {\it BeppoSAX} \citep{tagliaferri2003} and {\it XMM-Newton} \citep{foschini2006} provide evidence for a concave X-ray spectrum in the 0.1 -- 10~keV band, a signature of the presence of both the steep tail of the synchrotron emission and the rising part of IC spectrum. \par
EGRET on board the {\it Compton Gamma-ray Observatory (CGRO)} detected high-energy (HE, 0.1 GeV < E < 100~GeV) $\gamma$-ray emission from S5~0716+714 several times from 1991 to 1996 \citep{lin1995, hartman1999}. Two strong $\gamma$-ray flares in September and October 2007 were detected in the source with {\it AGILE} \citep{chen2008} in the HE range. These authors also carried out SED modeling of the source with two synchrotron self-Compton (SSC) emitting components, representative of slowly and rapidly variable components, respectively.\par
S5~0716+714 was first detected in the very-high-energy (VHE, E>100~GeV) range by MAGIC with  $5.8 \sigma$ significance level in November 2007 and then in April 2008 \citep{anderhub2009} during an optical flare. At that time MAGIC was working with a single telescope and the energy threshold of the telescope for such an high zenith range ($47^\circ<zd<55^\circ$) was 400 GeV. The analysis of multi-wavelength data suggested a correlation between the VHE $\gamma$-ray and optical emission. A structured jet model, composed of a fast spine surrounded by a slower moving layer \citep{ghisellini2005,tavecchio2009} better described the data compared to a simple one-zone SSC model. This source is also among the bright blazars in the {\it Fermi}-LAT (Large Area Telescope) Bright AGN Sample (LBAS) \citep{abdo2010LBAS} and in the \textit{Fermi} third catalog \citep{3fglpaper} it is among the ones with the highest variability index.
The combined GeV -- TeV spectrum of the source might display an absorption-like feature in the 10 -- 100~GeV energy range \citep{senturk2013}.\par
Apart from the $\gamma$-ray/optical connection, which is to be expected in one-zone, single population leptonic models, another interesting feature of the broadband activity was reported in \cite{rani2013a}, in which the authors found the $\gamma$-ray/optical flux variations leading the radio variability by $\sim$~65 days. An orphan X-ray flare was detected in 2009 by \textit{Swift}-XRT but no VHE observations were available to check if there was a counterpart in the VHE range. The behavior of the source in the past years can be contextualized in the scenario of a shockwave propagating along a helical path in the blazar's jet \citep{marscher2008,larionov2013}.\par
Recently, more attention was given to the electric vector position angle swings: \cite{rani2014} found a significant correlation between the inner jet outflow orientation and $\gamma$-ray flux variations, showing how the morphology of the inner jet has a strong connection with the $\gamma$-ray flares. \cite{chandra2015} studied the rapid variation in the degree of polarization (PD) and in the polarization angle (PA) within the Helical Magnetic Field Model (HMFM) \citep{zhang2014}, explaining such features as most likely due to reconnections in the emission region of the jet.\par
In the present work, we report the results of a multi-wavelength (MWL) campaign organised to follow an unprecedented outburst phase of the blazar S5~0716+714 during January 2015. The source was detected at its historic highest brightness at optical and IR bands. On January 11, 2015, (MJD~57033) the NIR photometry reported an increase of its flux by a factor of 2.5 in the NIR band in a rather short lapse of 12 days \citep{atel6902,atel6962}. During the night of 18 January 2015 (MJD~57040), the source was detected at its historic high brightness, with R band magnitude $\sim$11.71 \citep{atel6957}.
The TeV observations triggered to follow the exceptionally high optical state detected the source at energies above 150 GeV \citep{atel6999} and went on until the flaring activity faded away. \par
The paper is structured as follows. In Section 2 and subsections we describe the various instruments involved in the observations and their data analysis. Then in Section 3 we present the multi-wavelength light curves and discuss the peculiarities of the source activity in the different bands. Section 4 is devoted to the jet analysis of the object by VLBI. In Section 5 the spectral fitting of \textit{NuSTAR} and \textit{Swift}-XRT data are shown. In Section 6 the VHE spectra obtained by MAGIC and the redshift estimation using MAGIC and \textit{Fermi}-LAT simultaneous data are presented, and in Section 7 the broadband SED for the two phases considered is discussed together with the modeling of the source. Results and Conclusions are given in Section 8. 

\section{Observations and analysis}
\subsection{VHE $\gamma$-ray observations}
MAGIC is a stereoscopic system consisting of two 17\,m diameter Imaging Atmospheric Cherenkov Telescopes located at the Observatorio del Roque de los Muchachos, on the Canary Island of La Palma. The current sensitivity for medium-zenith observations ($30^\circ<zd<45^\circ$) above 210\,GeV is $0.76 \pm 0.04\,\%$ of the Crab Nebula's flux in 50~h \citep{Aleksic2016_b}. On 19 January 2015 (MJD~57041), triggered by the high optical state and by high-energy photons detected by {\it Fermi}-LAT, MAGIC started to observe S5~0716+714. On 23 January 2015 (MJD~57045) the significance of the signal was found to be 6.4 $\sigma$, and it reached a maximum value of 13.2 $\sigma$ on 26 January (MJD~57048): the flux increased from $(4.1 \pm 1.1) \times 10^{-11} \textrm{cm}^{-2} \ \textrm{s}^{-1}$ to $(8.9 \pm 1.1) \times 10^{-11} \textrm{cm}^{-2} \ \textrm{s}^{-1}$ above 150 GeV, which is the highest level ever detected in the VHE band for this source. The next activity of S5~0716+714 in the VHE range was detected by MAGIC on 13 February (MJD~57066), this time lasting four days only, up to the 16 February (MJD~57069). In the present work the two multi-wavelength periods of observations that include MAGIC data are indicated as Phase A, from 18 to 27 January 2015 (MJD~57040 to MJD~57050), and Phase B from 12 to 17 February 2015 (MJD~57065 to MJD~57070). We collected 17.74 hours of data in the zenith angle range of $40^\circ<zd<50^\circ$ and the analysis was performed using the standard MAGIC analysis framework MARS \citep{zanin2013, Aleksic2016_b}. After the applied quality cuts, the survived events amount to 17.46 hours in total. A statistical significance of 18.9 $\sigma$ was found for the full sample after cuts. The significance of the signal was calculated as in Eq. 17 in \cite{lima1983}. For the MAGIC SED, the systematic uncertainty on the flux normalization was estimated to be 11\%, and on the spectral slope $\pm$ 0.15. The analysis energy threshold is  $\sim$~125 GeV, measured as the peak of the Monte Carlo (MC) energy distribution for a source with the spectral shape of S5~0716+714. A full description of the MAGIC systematic uncertainties can be found in \cite{Aleksic2016_b}.

\subsection{HE $\gamma$-ray observations}
The HE $\gamma$-ray (0.1--300~GeV) observations for a time period between 1 November 2014 (MJD~56962) and 31 July 2015 (MJD~57234) were obtained in a survey mode by the {\it Fermi}-LAT \citep{atwood2009}. The LAT data were analyzed using the standard ScienceTools\footnote{https://fermi.gsfc.nasa.gov/ssc/data/analysis/} (software version v10.01.01, pass8) with instrument response function P8R2\_SOURCE\_V6. Photons in the source event class were selected for the analysis. We analyzed a region of interest (ROI) of 10$^{\circ}$ in radius centered at the position of S5~0716+714 using a maximum-likelihood algorithm \citep{mattox1996}. We included all 54 sources of the 3FGL catalog \citep{3fglpaper} within 20$^{\circ}$ of the position of S5~0716+714 in the unbinned likelihood analysis. Model parameters for sources within 5$^\circ$ of the ROI are kept free; we kept the model parameters for the rest fixed to their catalogue values.\par
To investigate the source variability at E~>100~MeV, we generated the daily-binned photon flux and index curves using unbinned likelihood analysis. The daily binned data were computed by modeling the spectra by a  power-law model (PWL, N(E) = N$_0$ E$^{-\Gamma}$, N$_0$ : prefactor, and $\Gamma$ : power law index). We examined the spectral behavior over the whole energy range with a PWL model fitting over equally spaced logarithmic energy bins with $\Gamma$ kept constant and equal to the value fit over the whole range. Even if in the \textit{Fermi} third catalog \citep{3fglpaper} the source spectrum is described by a log parabola model, for the present dedicated analysis the PWL model better fits the data. The upper limits for the light curve are shown as grey triangles in the second panel from top in Fig.~\ref{fig:MWL_LC} and they were calculated for test statistics <9.

\subsection{X-ray and optical/UV observations}
\subsubsection{\textit{NuSTAR}}
The exceptional flare from the object triggered a Target of Opportunity observation of the object by \textit{NuSTAR}. \textit{NuSTAR}, a NASA Small Explorer satellite sensitive in the hard X-ray band features two multilayer-coated telescopes, focusing the reflected X-rays on the pixellated CdZnTe focal plane modules with the half-power diameter of an image of a point source of $\sim 1 '$. It provides a bandpass of 3-79 keV with spectral resolution of $\sim 1$ keV. For more details, see \cite{harrison2013}. \par
After screening for the South Atlantic Anomaly passages and Earth occultation, the pointing (with OBSID 90002003002) resulted in roughly 14.9~ks of net observing time on 24 January 2015 (MJD~57046). After processing the the raw data with the \textit{NuSTAR} Data Analysis Software (NuSTARDAS) package v.1.3.1 (via the script \texttt{nupipeline}), the source data were extracted from a region of $45''$ radius, centered on the centroid of X-ray emission, while the background was extracted from a $1.5'$ radius region roughly $5'$ SW of the source location. Spectra were binned in order to have at least 30 counts per rebinned channel.  We considered the spectral channels corresponding nominally to the 3-60 keV energy range, where the source was robustly detected. The mean net (background-subtracted) count rates were $0.174 \pm 0.003$ and $0.165 \pm 0.003$  cts s$^{-1}$, respectively, for the modules FPMA and FPMB. We found no variability of the source as a function of time within the \textit{NuSTAR} observation, and we summed the data into one spectral file for each focal plane module.

\subsubsection{\textit{Swift}-XRT}
The multi epochs (35) event-list obtained by the X-ray Telescope (\textit{XRT}) \citep{burrows2004}, on-board the \textit{Neil Gehrels Swift} satellite in the period 1 January 2015 (MJD~57023.2) to 28 February 2015 (MJD~57081.2) with total exposure time of $\sim$16.81 hours were downloaded from publicly available SWIFTXRLOG (\textit{Swift} XRT Instrument Log). They were processed using the procedure described in \cite{ramazani2017}. All these observations have been performed in photon counting (PC) mode, with an average integration time of 1.7~ks each. The equivalent Galactic hydrogen column density is fixed to the value of $n_H = 3.11 \times10^{20}\textrm{cm}^{-2}$ \citep{kalberla2005}. We performed spectral fits to all 35 epochs using a simple power law model, with Galactic absorption. This model provides a good fit, and the 0.3 -10~keV spectral index is ${2.72} \pm {0.1}$. We discuss the Swift observations more extensively in Section 3. We note specifically that one of the pointings was simultaneous with the \textit{NuSTAR} observation (at MJD~57045--24 January 2015) and we use that observation for joint XRT - \textit{NuSTAR} spectral fitting in Section~\ref{sec:nustar-swift}.

\subsubsection{\textit{Swift}-UVOT}
Photometric observations by the \textit{Swift} Ultra-Violet and Optical Telescope (UVOT) instrument were made in the three UV (uvw2, uvm2, and uvw1) and three optical (u, b, and v) filters in both imaging and event mode. During the February 2015 observations, good coverage in the UV and u bands exist and in particular uvm2 data in event mode were obtained in order to resolve short timescale variability in the UV.\par
The UVOT data reduction used the Heasarc Heasoft version 6.16 and \textit{Swift} CALDB (September 2013). The event mode data were typically 800~s long exposures and were binned into shorter time slices, converted into images, then aspect corrected.\par
This analysis allowed the identification of a few observations for which the pointing drift was too large, and those were excluded from further analysis. Data taken in image mode were similarly validated. The magnitudes were determined from the images using the UVOTMAGHIST program using the standard calibration \citep{poole2008, breeveld2011}. The details of the event mode data processing are as follows: GTI extensions were created in the event file for the desired time intervals; UVOTATTJUMPCORR was run to improve the attitude file; COORDINATOR and UVOTSCREEN were used to correct the event file, which was then processed using UVOTIMAGE so the individual data could be inspected.

\subsection{Optical observations and polarimetry data}
The Tuorla Blazar monitoring program\footnote{http://users.utu.fi/kani/1m} collects blazar optical light curves in the R band from several observatories. The present work shows in particular data from the 1.03~m telescope at Tuorla Observatory, Finland and the 35~cm telescope at the KVA observatory on La Palma, Canary Islands, Spain. The data are analyzed with a semi-automatic pipeline using standard procedures \citep{nilsson2017}\par
The Boston University (BU) group uses the 1.83~m Perkins Telescope at Lowell Observatory (Flagstaff, AZ) to carry out optical observations of a sample of $\gamma$-ray blazars, including S5~0716+714. The telescope is equipped with the PRISM camera operating in photometric (UBVRI) and polarimetric modes. The details of the observations and data reduction can be found in \cite{jorstad2010}.\par
The optical polarimetric data presented here are complemented by data from the RINGO3 polarimeter on the 2.0~m fully robotic Liverpool Telescope on the Roque the los Muchachos Observatory (La Palma, Canary Islands, Spain). The RINGO3 data presented here were obtained as part of a monitoring program of bright optical blazars \citep{jermak2016}. S5~0716+714 is one of the targets regularly monitored by this program, with a time cadence of $\sim~3$ days. The RINGO3 polarimeter acquires polarimetric measurements in three different passbands recorded in the so called "Red", "Green", and "Blue" cameras\footnote{See the RINGO3 specifications at:\\ http://telescope.livjm.ac.uk/TelInst/Inst/RINGO3/}. Here we only present the data from the "Green" camera (the one with the closest wavelength passband to R-band). The RINGO3 data were reduced following the procedure explained in \cite{steele2017}.\par
R-band photometry and polarimetry observations of S5~0716+714 were performed using the HONIR (Hiroshima Optical and Near-InfraRed camera) instrument installed on the 1.5~m Kanata telescope located at the Higashi-Hiroshima Observatory, Japan \citep{akitaya2014}. A sequence of photopolarimetric observations consisted of successive exposures at 4 position angles of a half-wave plate: 0, 45, 22.5 and 67.5~deg. The data were reduced under the standard procedure of CCD photometry. The aperture photometry was performed using the APPHOT package in PYRAF\footnote{http://www.stsci.edu/institute/software\_hardware/pyraf/}, and the differential photometry with a comparison star taken in the same frame of S5~0716+714. The comparison star is located at R.A. = 07:21:53.44 and Decl. = +71:20:36.0 (J2000), and its magnitude is R = 14.032 (UCAC-4 Catalog). The polarization angle is defined as usual (measured from north to east), based on calibrations with polarized stars, HD183143 and HD204827 \citep{schulz1983}. The systematic error caused by instrumental polarization was smaller than 2~deg using the polarized stars.\par
St.~Petersburg University optical photometric and polarimetric data are from the 70~cm AZT-8 telescope of the Crimean Astrophysical Observatory\footnote{http://craocrimea.ru/ru} and the 40~cm telescope LX-200 in St.~Petersburg, both equipped with nearly identical imaging photometers-polarimeters. Polarimetric observations were performed using two Savart plates rotated by 45\degr~relative to each other \citep[see][]{larionov2008b}. Instrumental polarization was found via stars located near the object under the assumption that their radiation is unpolarized. This is indicated also by the low level of extinction in the direction of S5~0716+714 ($A_V=0.085$ mag; $A_R=0.067$ mag; \citealt{schlafly2011}).\par

\subsection{Radio observations}
\label{sec:radio_observations}
The 230~GHz (1.33~mm) flux density data were obtained at the Submillimeter Array (SMA) near the summit of Mauna Kea (Hawaii). S5~0716+714 is included in an ongoing monitoring program at the SMA to determine the flux densities of compact extragalactic radio sources that can be used as complex gain calibrators at mm wavelengths \citep{gurwell2007}. Potential calibrators are from time to time observed for 3 to 5 minutes. Data from this program are updated regularly and are available at the SMA website\footnote{http://sma1.sma.hawaii.edu/callist/callist.html}, while the present analysis was a dedicated one.\par
The IRAM 30~m millimeter Radiotelescope provided 230~GHz (1.3~mm) and 86~GHz (3.5~mm) data that were obtained as part of the POLAMI\footnote{http://polami.iaa.es} (Polarimetric AGN Monitoring at Millimeter Wavelengths) program, see \cite{agudo2018a,agudo2018b} and \cite{thum2018}. The POLAMI data of S5~0716+714 were acquired and reduced as described in detail in \cite{agudo2018a}. The CARMA data were taken with the eight 3.5\,m antennas as part of the Monitoring of $\gamma$-ray Active galactic nuclei with Radio, Millimeter and Optical Telescopes (MARMOT) project\footnote{http://www.astro.caltech.edu/marmot/}. We used 7.5\,GHz of bandwidth with a center frequency of 94\,GHz. The integration time on S5~0716+714 was 5 minutes for each observation, which yields a typical rms of $10-110$\,mJy. Absolute flux density calibration was done using nearby observations of Mars. The observational errors are dominated by the absolute calibration uncertainty, assumed to be 10\%. All data were processed using the Multichannel Image Reconstruction Image Analysis and Display \citep[MIRIAD;][]{sault1995}.\par
We have analyzed Very Long Baseline Array (VLBA) data obtained for S5~0716+714, within the VLBA-BU-Blazar program\footnote{www.bu.edu/blazars/VLBAproject.html}, which are contemporaneous with the high energy event in January and February 2015. The data include total and polarized intensity images at 43~GHz at 9 epochs from November 2014 to August 2015. They were reduced using the Astronomical Image Processing System (AIPS)\footnote{http://www.aips.nrao.edu/index.shtml} and {\it Difmap}\footnote{ftp://ftp.eso.org/scisoft/scisoft4/sources/difmap/difmap.html} software packages, in the general manner described by \cite{jorstad2017,jorstad2005}. The total intensity images were modeled by components with circular Gaussian brightness distributions. This allows us to determine the minimum number of components that provides the best fit between  the data and model at each epoch, as well as the following parameters of components: flux density, $S$, distance from the core, $r$, position angle with respect to the core, $\Theta$, and size of the component, $a$ (FWHM of the Gaussian). These parameters are given in Table~\ref{table:ParKnots} of Appendix \ref{sec:appendix}.\par
The 37~GHz observations were made with the Mets\"ahovi radio telescope. The measurements were made with a 1~GHz-band dual beam receiver centered at 36.8~GHz. The observations are ON--ON observations, alternating the source and the sky in each feed horn. A typical integration time to obtain one flux density data point is between 1200 and 1800~s. Data points with a signal-to-noise ratio < 4 are handled as non-detections. The flux density scale is set by observations of DR~21. A detailed description of the data reduction and analysis is given in \cite{terasranta1998}. The error estimate in the flux density includes the contribution from the measurement rms and the uncertainty of the absolute calibration.\par
S5~0716+714 was observed at 15~GHz as part of a high-cadence $\gamma$-ray blazar monitoring program using the Owens Valley Radio Observatory (OVRO) 40~m telescope \citep{richards2011}. The OVRO 40~m uses off-axis dual-beam optics and a cryogenic pseudo-correlation receiver with a 15.0~GHz center frequency and 3~GHz bandwidth. The source is alternated between the two beams in an ON-ON fashion to remove atmospheric and ground contamination. The fast gain variations are corrected using a 180 degree phase switch. Calibration is achieved using a temperature-stable diode noise source to remove receiver gain drifts, and the flux density scale is derived assuming the value of 3.44~Jy at 15.0~GHz in \cite{baars1977}. The systematic uncertainty of about 5\% in the flux density scale is not included in the error bars. Complete details of the reduction and calibration procedure are found in \cite{richards2011}.\par
Observations at 2.6, 4.8, 10, and 15~GHz radio bands were conducted using the Effelsberg 100~m radio telescope\footnote{http://www.mpifr-bonn.mpg.de/en/effelsberg}. Measurements for the target source and for the calibrator sources
were made quasi-simultaneously using the cross-scan method slewing over the source position, in azimuth and elevation direction in order to gain the desired sensitivity. Subsequently, atmospheric opacity correction, pointing off-set correction, gain correction, and sensitivity correction were applied to the data. Details of the observations and data reduction are referred to \cite{angelakis2015}.\par 
The sources 3C~286, NGC~7027 and 3C~84 have been used as common calibrators for the instruments listed in this section.\par

\section{Multi-wavelength light curves}
\begin{figure*}[htbp!]
   \resizebox{\hsize}{!}
            {\includegraphics[]{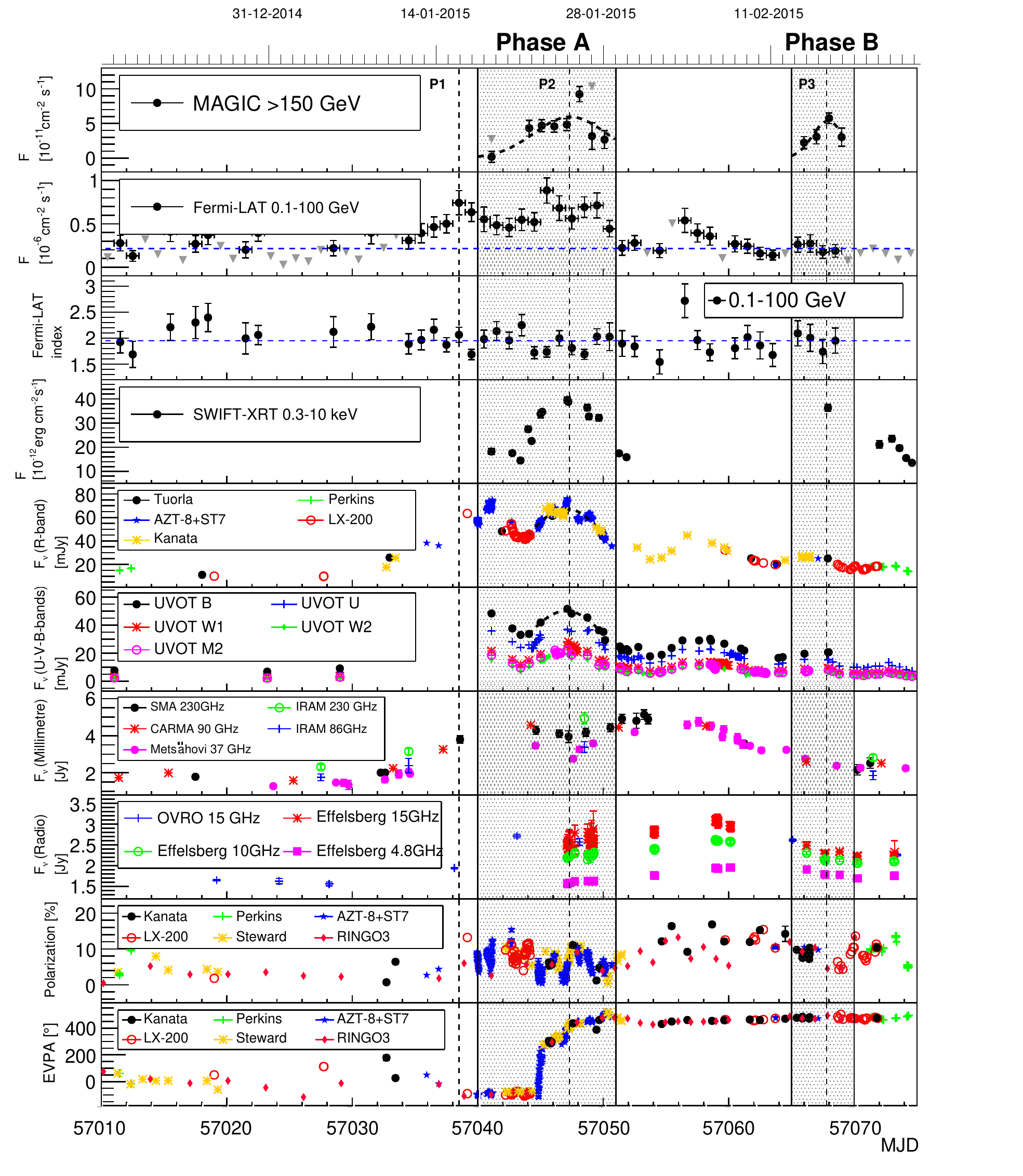}}
      \caption{Multi-wavelength flux and index curves of S5~0716+714 during the period from MJD~57010 to MJD~57080 (19 December 2014 to 27 February 2015). The shadowed areas indicate Phase A (from 18 to 27 January 2015 -- MJD~57040 to MJD~57050) and Phase B (from 12 to 17 February 2015 -- MJD~57065 to MJD~57070) high states in the VHE range and the corresponding activity in the other bands. P1, P2 and P3 (vertical dashed lines) indicate peaks in the HE and VHE emission.}
         \label{fig:MWL_LC}
    \end{figure*}

In Fig.~\ref{fig:MWL_LC} we present the multi-wavelength data collected during the course of the campaign. A summary of the most important dates can be found in Table~\ref{table:dates}.
\begin{table*}[!htbp]
\caption{Summary of important dates}             % title of Table
\label{table:dates}      % is used to refer this table in the text
\centering                          % used for centering table
\begin{tabular}{c c l}        % centered columns (4 columns)
\hline\hline                 % inserts double horizontal lines
MJD & Calendar date & Description\\
\hline
57038.5                           &       16 January 2015       &       P1: first peak of the HE emission $\rightarrow$ trigger VHE observations          \\
57040           &       18 January 2015       &       start of Phase A       \\
 %               &                               &                                                       \\
57044/45	&	22/23 January 2015	&	1day EVPA rotation of $\sim$360$^\circ$	        \\
57047.3 $\pm$ 0.53                &       25 January 2015       &       P2: Gaussian fit peak of the VHE emission in PHASE A                 \\
57050           &       28 January 2015       &       end of Phase A                                  \\
57050 $\pm$ 3		&       28 January 2015	&	K14b passage through A1     	    	     	      			\\
57056           &       03 February 2015      &    R4: Gaussian fit peak of radio emission in the intermediate phase \\
57065           &       12 February 2015      &       start of Phase B                                \\
57067.8 $\pm$ 0.23		&	14 February 2015	&	P3: Gaussian fit peak of the VHE emission in Phase B                 \\
%                &                               &                                                       \\
57070           &       16 February 2015      &       end of Phase B         \\
57092           &       11 March 2015      &    R5: Gaussian fit peak of radio emission        \\
\hline                                   %inserts single line
\end{tabular}
\end{table*}
    \begin{figure*}[htbp!]
\centering
    \includegraphics[width=0.60\textwidth]{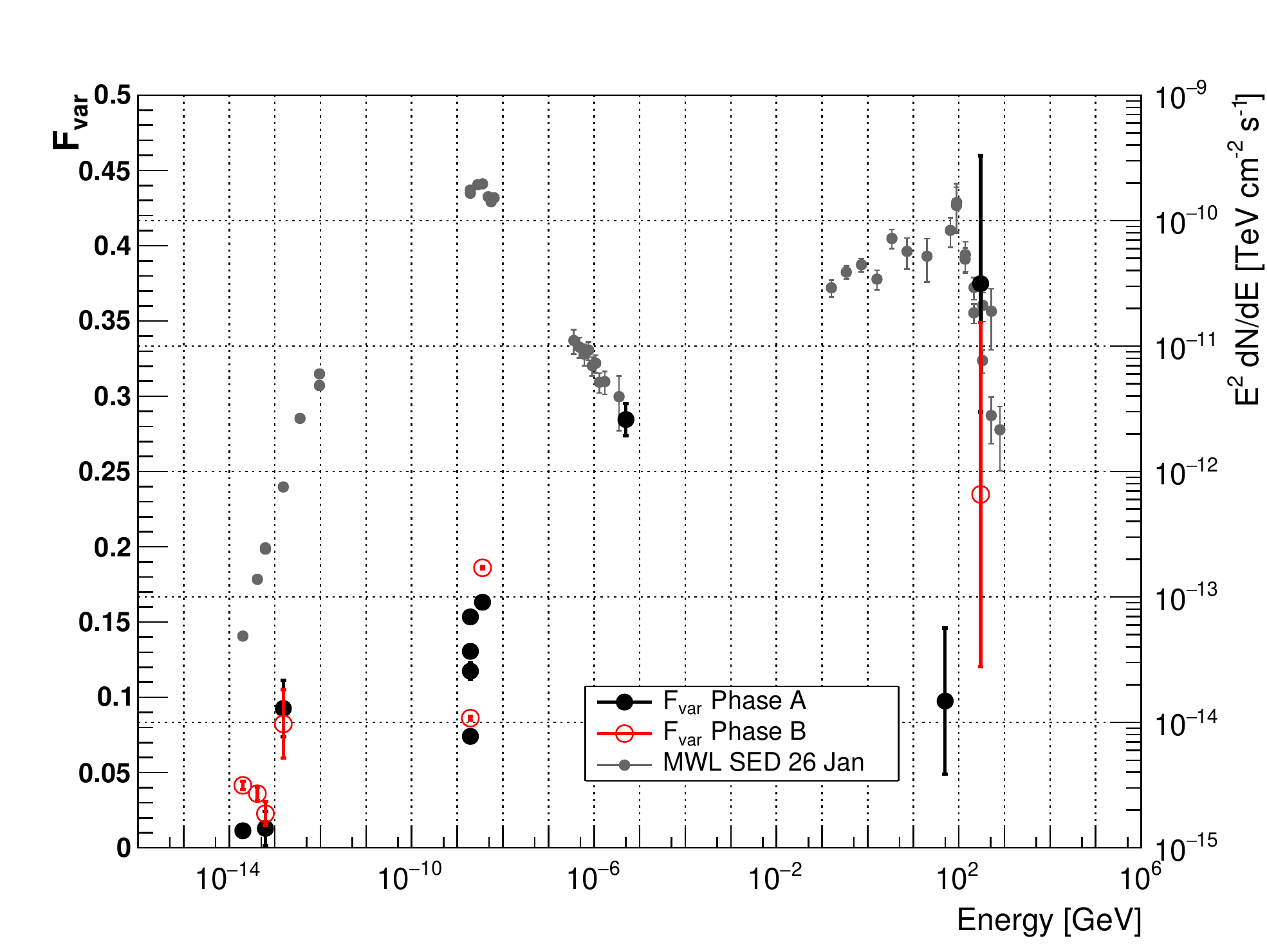}  
    \caption{Fractional variability ($F_\text{var}$) as a function of the energy for Phase A and Phase B. Vertical bars denote 1$\sigma$ uncertainties. In gray a Spectral Energy Distribution of the source is overlayed (a snapshot of 26 January 2015), to make easier to associate the value of the fractional variability to the corresponding energy band.}
    \label{fig:f_var}
%\end{figure}
\end{figure*} 
The top panel shows the daily-binned MAGIC light curve: in the VHE band, the no-variability hypothesis has been discarded, since the fit for a constant flux resulted in a $\chi^2/n.d.f.=42/7$ for Phase A and  $\chi^2/n.d.f.=10/3$ for Phase B data. A Gaussian function better fits the flare shape, providing $\chi^2/n.d.f.=15/5$ and $\chi^2/n.d.f.=1.8/1$ for the two phases respectively. 
Phase A peaks on 25 January, MJD 57047.3 $\pm 0.4$ (in Fig.~\ref{fig:MWL_LC} corresponding to the P2 vertical dashed line) with a flux of $(5.9 \pm 0.5)\times10^{-11}\textrm{cm}^{-2} \ \textrm{s}^{-1}$; the standard deviation of the fit is $\sigma =(2.8 \pm 0.5)$ days. Phase B peaks on 14 February 2015 (MJD~57067.9 $\pm 0.2$) with a corresponding flux of $(5.3 \pm 0.7)\times10^{-11} \textrm{cm}^{-2} \ \textrm{s}^{-1}$ indicated by the vertical dashed line P4. The standard deviation of the fit is $\sigma =(1.2 \pm 0.3)$ days. Intra-day variability (in shorter time scales with respect to the daily-binned light curves) was not detected with MAGIC: the light curve was fit at different time intervals down to 5 minutes, but a constant fit was found to be consistent with the data up to the daily scale, where variability is significant. \par
The \textit{Fermi}-LAT daily-binned light curve is shown in the second panel from top: the first peak visible in the curve, marked with the vertical dashed line P1, is the precursor of the whole flaring activity. After P1, (MJD~57038.5, 16 January 2017), which triggered VHE observations, two other peaks are visible in Phase A, and the maximum flux reached in HE is $(8.8 \pm 1.4) \times10^{-7} \textrm{cm}^{-2} \ \textrm{s}^{-1}$, 4 times the average flux in HE for the source from the \textit{Fermi}-LAT 3FGL Catalog \citep{3fglpaper}. The photon index for \textit{Fermi}-LAT observations stays very close to the average \textit{Fermi}-LAT index of the source, indicated in the corresponding panel of Fig.~\ref{fig:MWL_LC} with the horizontal dashed blue line. \par
The average integral photon X-ray flux (0.3--10~keV) reported by \textit{Swift}-XRT in Fig.~\ref{fig:MWL_LC}, fourth panel from top, is $(1.87 \pm 0.46) \times10^{-11} \textrm{erg} \ \textrm{cm}^{-2} \ \textrm{s}^{-1}$. The X-ray flux peaking at MJD~57047 (25 January 2015) with $F_{(0.3-10 \textrm{keV})} = (3.95 \pm 0.12) \times10^{-11} \textrm{erg} \ \textrm{cm}^{-2} \ \textrm{s}^{-1}$ which is a factor of $\sim$~2 higher than the average flux of the analysed period. The X-ray spectral index during the analysed period varies between $2.03 \le \Gamma_{X} \le 2.56$. The softest spectral index was obtained on the night of the highest X-ray and VHE $\gamma$-ray flux (MJD~57047--25 January 2015). The X-ray spectra start hardening smoothly afterwards for the next 14 consecutive nights. The X-ray spectra on the nights of VHE $\gamma$-ray flares of Phase A and B can be well described by a power law with spectral index of $\Gamma_{X, MJD~57047} = 2.56 \pm 0.06$ (reduced $\chi^{2} / \textrm{n.d.f.}=0.738/29$ ) and $\Gamma_{X, MJD~57067} = 2.33 \pm 0.06$ (reduced $\chi^{2} / \textrm{n.d.f.} = 0.871/26$) respectively.\par
From the Tuorla optical monitoring, on 18 January 2015 (MJD~57040) the magnitude in the R-band reached a value of $\sim$~11.5, higher than the magnitude value of the source in the quiescent state (R$_{mag} \sim$~13.5). Later, the source faded down to R$_{mag}$ = 12.18 $\pm$ 0.03 on 20 January 2015 (MJD~57042), flared up again to R$_{mag}$ = 11.77 $\pm$ 0.02 on 23 January (MJD~57045), and then demonstrated rapid flux variations above R$_{mag}$ = 11.9 \citep{atel7004}.
In general, the optical light curves and the \textit{Swift}-UVOT light curves show the same trend, a double peaked shape with the second peak coincident with the dashed vertical line P2 in Fig.~\ref{fig:MWL_LC}. P2 is in fact identifying a peak not only in the VHE light curve, but also in the X-ray, optical, and UV bands. AZT-8+ST7 light curve can be fit by a Gaussian peaking on MJD~57047 (25 January 2015) at a flux of $(67.49 \pm 0.01)$ mJy. The standard deviation of the fit is $\sigma =3.366 \pm 0.001$ days. The \textit{Swift}-UVOT light curve can be fit by a Gaussian peaking on MJD~57047 (25 January 2015) at a flux of $(50.2 \pm 0.1)$~mJy. The standard deviation of the fit is $\sigma =3.36 \pm 0.02$ days.\par
The activity in the radio band shown in Fig.~\ref{fig:MWL_LC} in Phase A and Phase B is moderate compared to the other energy bands but does not describe a simply quiescent radio state. In the past the source has gone through many high states in the radio band, for instance the one described in \cite{rani2013a}, Fig.~2, named R6. During the R6 flare, the highest flux density reported was $\sim$ 10~Jy while in the present one, the highest level of radio flux density at the same frequency (230~GHz) is only 5~Jy. We study in more detail the possible delay between radio and optical/$\gamma$-ray bands in Sec.~\ref{sec:radio_delay}.\par
As reported in \cite{chandra2015}, the Phase A flare presents a double peaked shape in the HE $\gamma$-ray and optical bands. The feature is particularly evident in the R-band. The $\gamma$-ray light curve has the first sub-peak (in Fig.~\ref{fig:MWL_LC} corresponding to the P1 vertical dashed line) located immediately before Phase A at MJD~57038.5 (16 January 2015). That indicates the optical/$\gamma$-ray emission as possible precursors of the VHE activity, whose peak starts to rise after MJD~57040 (18 January 2015) as indicated by the Gaussian fit of the VHE light curve in the top panel of Fig.~\ref{fig:MWL_LC}. \par
The Phase B flare is very different, being clearly visible in the VHE and X-ray band only. All the other bands are in a quiescent level, perhaps reproducing the conditions of the X-ray flare in December 2009 in \cite{rani2013a}, where VHE data were not available. In that case, the X-ray emission was described by both synchrotron and inverse Compton mechanisms in a single-zone, one-population leptonic model.\par
The fractional variability $F_{var}$ has been calculated using equation 10 in \cite{vaughan2003}:
\begin{equation}
F_\text{var} = \sqrt{\dfrac{S^2 - \overline{\sigma^2_\text{err}}}{\overline{x}^2}},
\end{equation}
which represents the normalized excess variance. $S$ stands for the standard deviation and $\overline{\sigma^2_\text{err}}$ the mean square error of the flux measurements, while $\bar{x}$ indicates the average flux. The uncertainty of $F_\text{var}$ is given by Eq.(2) in \cite{Aleksic2015}, after \cite{poutanen2008}. $F_\text{var}$ was calculated for all the light curves shown in Fig.~\ref{fig:MWL_LC} and the results are plotted in Fig.~\ref{fig:f_var} for both Phase A (full black dots) and Phase B (red open circles).\par
To make a direct comparison of the variability determined for the various energy bands, we  computed $F_\text{var}$ using only the multi-instrument observations strictly simultaneous to those performed by MAGIC.\par
\begin{figure*}[htbp!]
\centering
            {\includegraphics[width=0.90\textwidth]{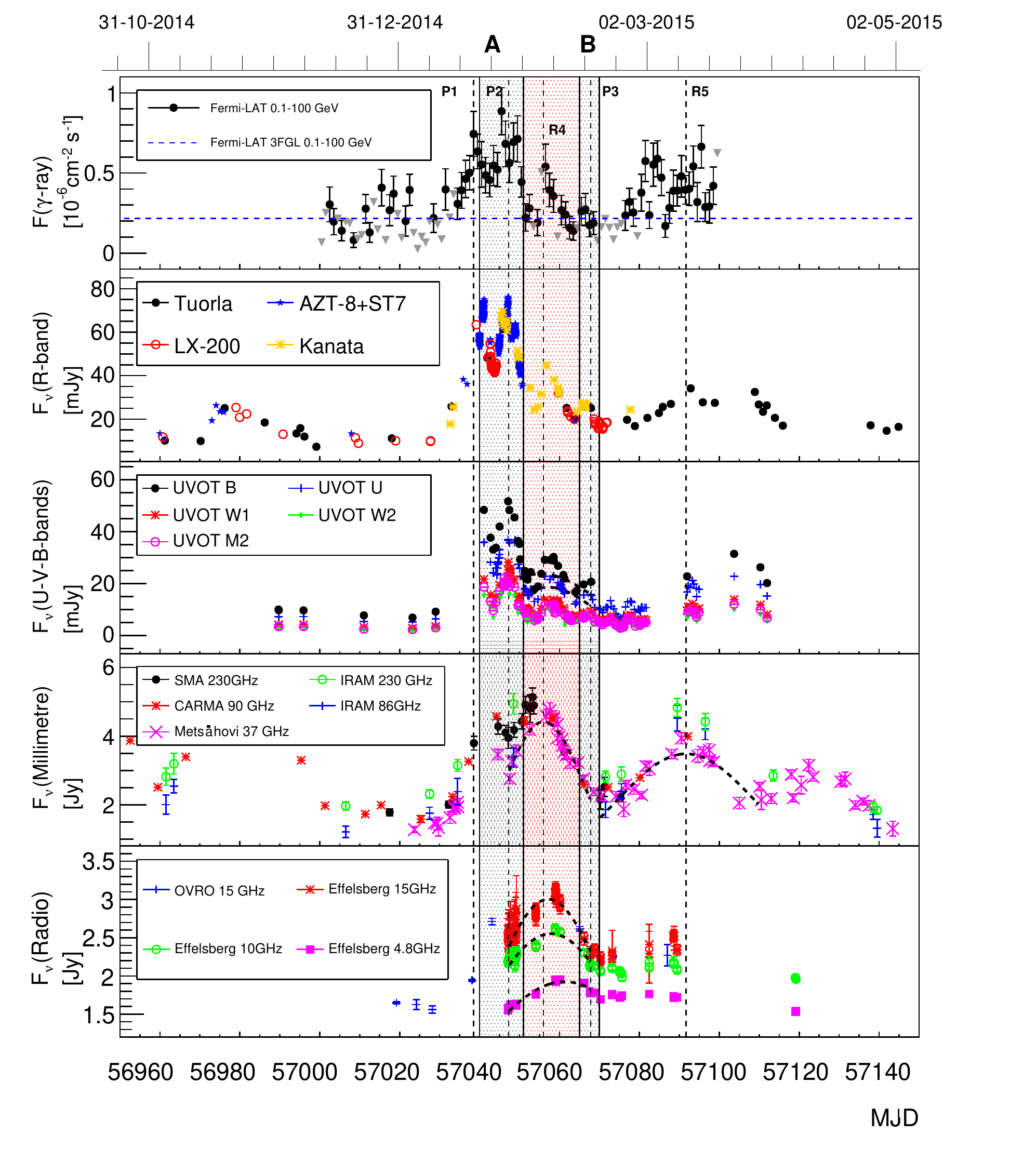}}
      \caption{HE $\gamma$-ray, optical and radio activity of S5~0716+714 during the period from MJD~56950 to MJD~57150 (20 October 2014 to 08 May 2015). The three shadowed areas indicate (from the left side) Phase A (from 18 to 27 January 2015 -- MJD~57040 to MJD~57050), an intermediate phase (from 28 January to 11 February 2015 -- MJD~57050 to MJD~57064), and Phase B (from 12 to 17 February 2015 -- MJD~57065 to MJD~57070). Vertical dashed lines mark important dates, as shown in Table~\ref{table:dates}.}
         \label{fig:radio}
    \end{figure*}
The overall behaviour of the fractional variability shows a rising tendency with increasing energy, at least up to the X-ray frequency. Since $F_\text{var}$ is highly sensitive to the sampling of the observed data, in the HE and VHE bands where the sampling is poorer, the results are affected by very large error bars, making impossible to confirm a general trend of $F_\text{var}$ for the whole energy spectrum. The highest fractional variabilities in Phase A occur in the VHE $\gamma$-ray band and in the X-ray band, with MAGIC ($F_\text{var}$=0.37 $\pm$ 0.08), and \textit{Swift}-XRT ($F_\text{var}$=0.28 $\pm$ 0.010). In Phase B the highest fractional variability is again in the VHE range with MAGIC ($F_\text{var}$=0.23 $\pm$ 0.13), not far from the one obtained by UVOT telescope ($F_\text{var}$=0.18 $\pm$ 0.8$\times10^{-3}$). 
\subsection{Analysis of radio activity in a larger time-frame}
\label{sec:radio_delay}
When MAGIC detected S5~0716+714 for the first time in 2008 \citep{anderhub2009}, the radio band was in a quiescent level. Here we see an increased activity in the low radio frequencies, especially in the intermediate period between Phase A and B. This activity could be just an effect of a previous smaller flare in high energy delayed by months as it seems typical for this source when considering longer periods of observation \citep{rani2013a,rani2014}.\par 
The present work includes one month of data, from the beginning of Phase A to the end of Phase B (MJD~57040 to MJD~57070). To have a better understanding of the radio behavior of the source we gathered $\gamma$-ray, optical and radio data in Fig.~\ref{fig:radio}, for a longer time-period of 8 months centered on 28 January 2015 (MJD~57050). VHE emission was not detected outside of Phase A and Phase B, but the data from the available instruments in the 8-months time window make it possible to better investigate the radio response and to compare the present dataset with the scenarios described in other previous multi-wavelength flares.\par
In Fig.~\ref{fig:radio} Phase A and Phase B are still defined by grey shadowed areas, while the intermediate phase is filled in light red. From Fig.~\ref{fig:radio}, second panel from the bottom, the radio activity in the intermediate zone between Phase A and Phase B (from MJD~57051 to MJD~57065--29 January to 12 February 2015) could be fit by a Gaussian shape ($\chi^2/n.d.f.=34.9/14$ and standard deviation of $\sigma =9.7 \pm 0.3$ days), with a maximum flux density of $4.42 \pm 0.07$ Jy, corresponding to MJD~57056 (03 February 2015).\par
In the bottom panel of Fig.~\ref{fig:radio}, which shows the radio activity in lower frequencies from the Effelsberg telescope, a peak could be identified by a Gaussian fit of the data ($\chi^2/n.d.f.=28.7/47$ and standard deviation of $\sigma =14.5 \pm 0.5$ days), for a maximum flux density at 15 GHz of $3.00 \pm 0.02$ Jy, corresponding to MJD~57057 (04 February 2015). If we consider the dashed vertical line R4 as the position of the radio peak (using the value of MJD~57056--03 February 2015 retrieved from the Mets\"ahovi data which has a smaller error), we can see a delay with respect to the $\gamma$-ray/optical peak P2 of $\sim9$ days, which is considerably smaller than the delays of $\sim65$ days or more found in \cite{rani2014}. Another hint of delayed activity, indicated by the dashed vertical line labeled as R5, can be seen in Mets\"ahovi data, with a maximum flux density of $3.48 \pm 0.06$ Jy, corresponding to MJD~57092 (11 March 2015), fit by a Gaussian with $\chi^2/n.d.f.=84.3/13$ and standard deviation of $\sigma =17.3 \pm 0.9$ days. In the latter case, the delay from the P2 flare would be longer, $\sim45$ days.\par
Based on data from April 2007 to January 2011, a considerable delay of the radio, $\sim65$ days from the optical/$\gamma$ flare, was found in \cite{rani2013a}. In \cite{rani2014} using a dataset from August 2008 to September 2013, the highest peak in radio flux occurred $\sim82 \pm 32$ days after the $\gamma$-ray one. The delay in the radio emission from the optical/$\gamma$ ones is supporting a scenario in which the $\gamma$-ray emission is produced upstream of the core while the radio one has its origin in a shock in the jet, first appearing and evolving in the innermost, ultracompact VLBI core region and subsequently moving downstream the jet at parsec scales with apparent superluminal speeds. Similar results, on a larger sample of blazars, are presented in \cite{fuhrmann2014}. A longer-term study centered on the flaring activity reported here could be interesting for future investigations but is beyond the scope of the present work.\par
\subsection{Electric Vector Position Angle swing}
\begin{figure}[htbp!]
\centering
            {\includegraphics[width=0.50\textwidth]{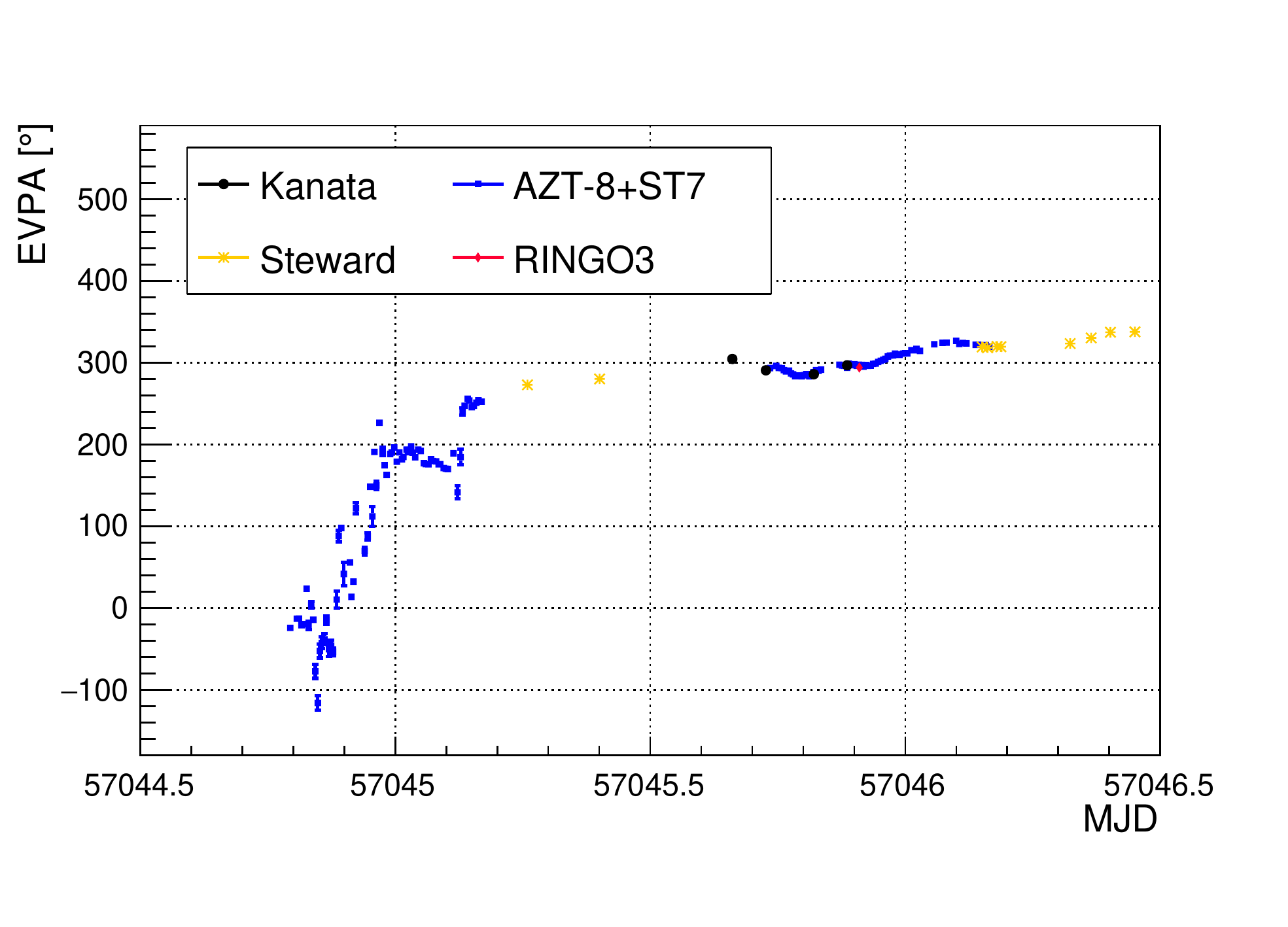}}
      \caption{Zoom of the Electric Vector Position Angle rotation of S5~0716+714 during the period from MJD~57043 to MJD~57048 (21 to 26 January 2015) in Phase A.}
         \label{fig:EVPA}
    \end{figure}

An important feature of Phase A is determined by the very fast change in the electric vector position angle (EVPA) happening over the night MJD~57044 (22 January 2015), 4 days after the first peak P1 in the optical band and 2 days before the MAGIC peak P2. This particular feature can be seen in the bottom panel of Fig.~\ref{fig:MWL_LC}, as well as in Fig.~\ref{fig:EVPA}, where the feature is zoomed in the time range MJD~57043-MJD~57048 (21-26 January 2015).\par
The dataset we presented put together EVPA data coming from many different instruments, as shown in the bottom panel of Fig.~\ref{fig:MWL_LC}. All the EVPA data collected are in agreement and were treated as in \cite{larionov2013}: in particular, to solve the $\pm 180^{\circ}$ ambiguity, we have added/subtracted $180^{\circ}$ each time that the subsequent value of EVPA was $>90^{\circ}$ less/more than the preceding one.\par
In \cite{marscher2008}, a similar behaviour of the EVPA was reported for BL Lacertae: a radio to $\gamma$-ray outburst, accompanied by a rotation of the EVPA, was observed as the consequence of a bright feature moving in the jet. In that case, the EVPA rotation was slower: it rotated steadily by about $240^{\circ}$ over a five-day interval before settling at a value of $195^{\circ}$.\par 
In 2008, when the source was observed in the VHE range for the first time \citep{anderhub2009}, the simultaneous optical outburst was accompanied by a $\sim360^{\circ}$ PA rotation of the electric vector, as reported in \cite{larionov2008a}. The rotation happened with an approximate rate of $60^{\circ}$ per day, so slower than in the present case. That rotation was interpreted as a consequence of the propagation of polarized emission from a knot spiraling down the jet. \cite{chandra2015} investigated the PA rotation swing from Phase A using the data from the Steward Observatory, in the frame of the HMFM \citep{zhang2014}, suggesting that the fast rotation was due to reconnections in the emission region in the jet.\par
\begin{figure*}[htbp!]
   \resizebox{\hsize}{!}
            {\includegraphics[]{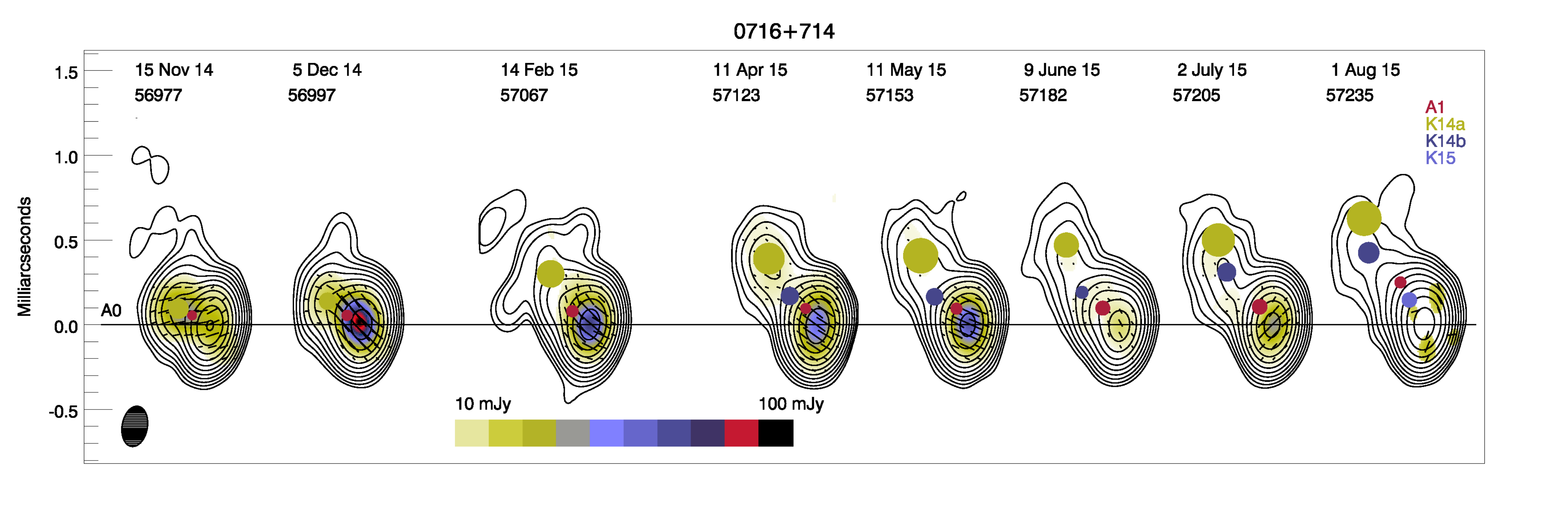}}
\caption{A sequence of total (contours) and polarized (color scale) intensity images of S5~0716+714 at 43~GHz, convolved with a beam of 0.24$\times$0.15 mas$^2$ at PA=-10$^\circ$. The global total intensity peak is 2655~mJy/beam and the global polarized intensity peak is 107~mJy/beam; black line segments within each image show the direction of polarization; the black horizontal line indicates the position of the core, A0.}
\label{fig:BU_vlba}
\end{figure*} 
\begin{figure*}[htbp!]
   \centering
   \begin{subfigure}[t]{8cm}
          \centering
          \includegraphics[width=8cm]{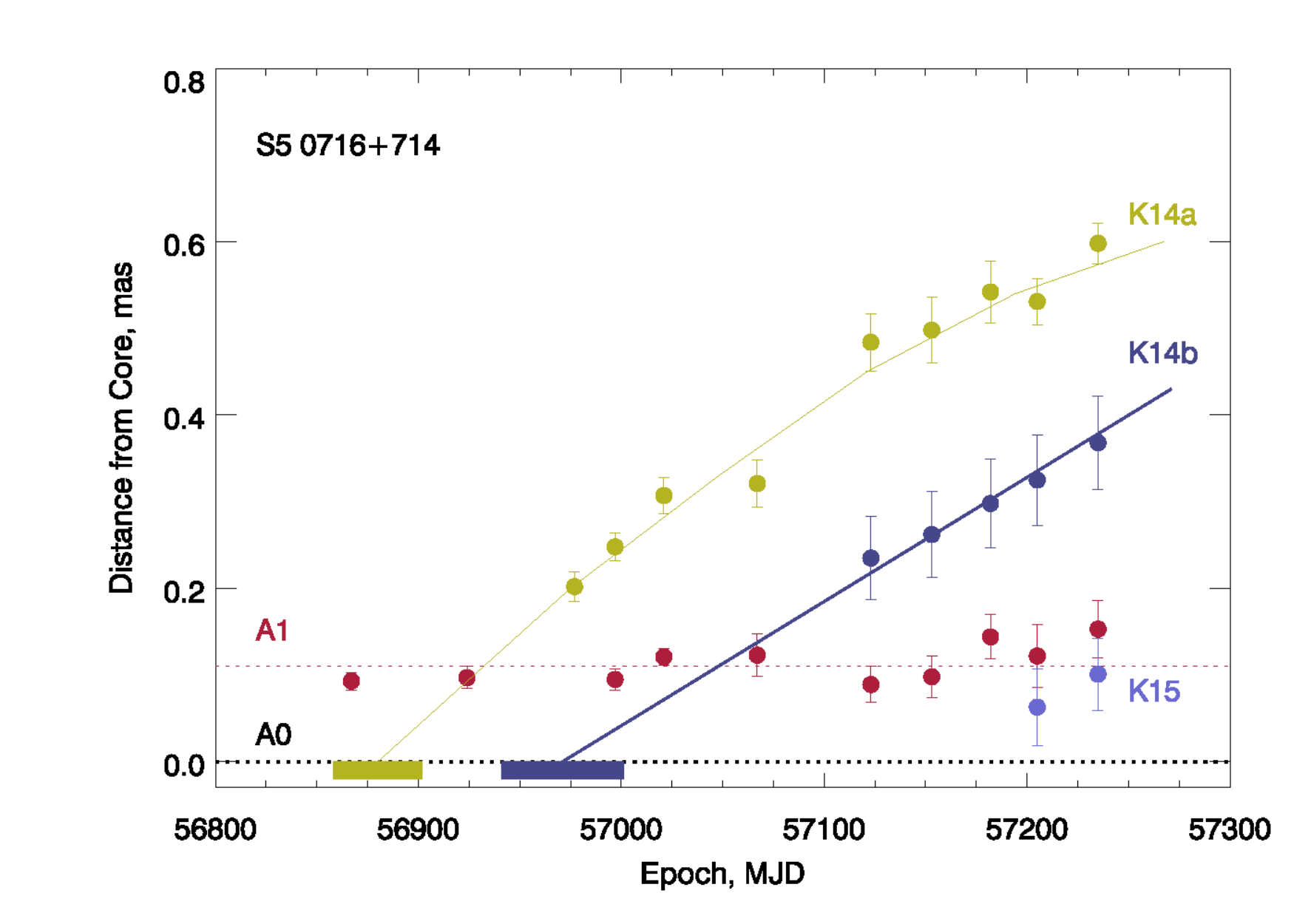}
          \caption{Separation of knots $A1$ (red), K14a (yellow), K14b (blue), and K15 (light blue) from the core $A0$ (black dotted line); the yellow and blue line segments on the position of $A0$ indicate the 1$\sigma$ uncertainty of the ejection times of K14a and K14b, respectively.
          \label{fig:BU_move}}
    \end{subfigure}
    \quad
    \begin{subfigure}[t]{8cm}
           \centering
           \includegraphics[width=8 cm]{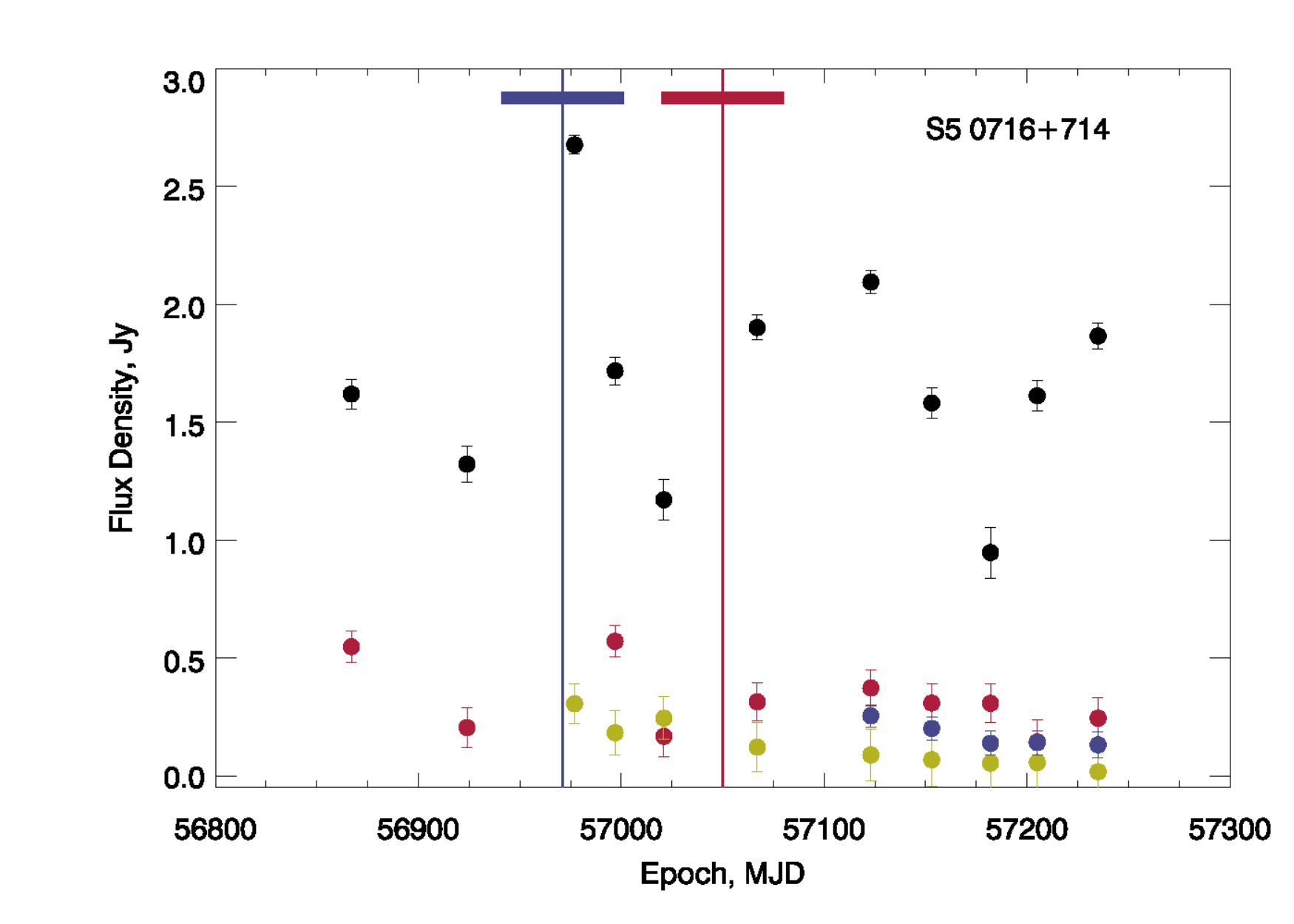}
           \caption{Light curves of the core A0 (black), stationary feature $A1$ (red), and moving knots K14a (yellow) and K14b (blue), flux densities of K14b are shifted by 0.1$\sim$Jy for clarity; vertical blue and red  lines indicate time of passage of K14b through $A0$ and $A1$, respectively. \label{fig:BU_lc}}
    \end{subfigure}
 \caption{VLBA-BU-BLAZAR analysis of S5~0716+714.}
 \label{fig:BU} 
\end{figure*}

\section{Jets evolution study with very-long-baseline interferometry}
\label{sec:VLBI}

We analyzed the structure of the jet with very-long-baseline interferometry. Details of the analysis have been presented in Section~\ref{sec:radio_observations}.\par
As shown in Fig.~\ref{fig:BU_vlba}, the core of the jet is the brightest knot, designated as $A0$, which is the southern-most feature of the jet, and assumed to be stationary. The position of the core across epochs is shown by the black horizontal line. In addition to the core, we have identified a stationary feature, $A1$, and two moving knots, K14a and K14b. Fig.~\ref{fig:BU_move} shows the evolution of the positions of the knots. Note that a stationary feature at a position similar to that of $A1$ was reported previously by \cite{rani2015} and \cite{jorstad2017} in data obtained from 2007 to 2013. The moving knots have apparent speeds of (10.7 $\pm$ 0.8)~c and (9.7 $\pm$ 1.8)~c for K14a and K14b, respectively, although K14a decelerates at $\sim$0.5~mas from the core. The direction of motion $\Phi$ is $\Phi_{K14a}$=(25.4$\pm$2.2) deg and $\Phi_{K14b}$=(43.3$\pm$5.6) deg respectively. Extrapolation of the motion of K14a and K14b back to the VLBI core suggests that K14a and K14b passed through the core on MJD~56880 $\pm$ 22 and MJD~56971 $\pm$ 30, respectively.\par
Knot K14b is of special interest with respect to the high energy event in January 2015 (MJD 57040-57050): according to its proper motion of (0.51 $\pm$ 0.09) mas/yr, K14b passed through A1 on MJD~57050 $\pm$ 30. This coincides with the high $\gamma$-ray state and TeV detection of S5~0716+714. Moreover, Fig.~\ref{fig:BU_lc} shows that $A1$ had an elevated flux density at the epoch closest to MJD~57050 when K14b should be passing through $A1$ according to its proper motion. In addition a change of the position angle of $A1$ from $\sim$80$^\circ$ to $\sim$50$^\circ$ is detected around the time of the expected passage (see Table~\ref{table:ParKnots}). The latter angle is close to the PA of K14b, $\Theta$=(47.5 $\pm$ 4.6) deg. This implies an interaction between superluminal knot K14b and stationary feature A1, and supports the hypothesis that $A1$ is a recollimation shock similar to that observed, e.g., in BL~Lacertae \citep{marscher2008, cohen2014}.\par
In addition, the average size of A1 is (0.049 $\pm$ 0.020)~mas, which implies that K14b needs (35 $\pm$ 13)~days to pass through the stationary feature. The latter agrees very well with the duration of 34 days of the elevated $\gamma$-ray flux in the \textit{Fermi} light curve of S5~0716+714, from MJD~57032 to 57066 (10 January 2015 to 13 February 2015). In this scenario the TeV detections can be associated with the entrance and exit of the superluminal knot in and out of the recollimation shock.
\section{Spectral fitting of \textit{NuSTAR} and \textit{Swift}-XRT data}
\label{sec:nustar-swift}
For \textit{NuSTAR} data, we performed the spectral fitting with XSPEC v12.8.2, with the standard instrumental response matrices and effective area files derived using the ftool {\tt nuproducts}. We fit the data for both \textit{NuSTAR} detectors simultaneously, allowing an offset of the normalization factor for module FPMB with respect to module FPMA.  Regardless of the adopted models, the normalization offset was less than 5\%. First, we adopted a simple power-law model modified by the effects of the Galactic absorption, corresponding to a column of $3.11 \times  10^{20}$ cm$^{-2}$ \citep{kalberla2005}. The fit returns the power-law index of $1.93 \pm 0.04$, but the residuals show that the \textit{NuSTAR} spectrum is more concave (i.e., the spectrum gets flatter towards higher energies) than a simple power-law model would imply. In addition, this index is significantly harder than that inferred from the \textit{Swift}-XRT data alone (which shows the index of $\sim~2.75$ ), which also suggests a more complex spectral model.\par
Since the \textit{NuSTAR} and \textit{Swift} data were nearly strictly contemporaneous (with a significant overlap) we fit the \textit{NuSTAR} and \textit{Swift}-XRT data simultaneously, but allowing for the normalizations of \textit{Swift} and \textit{NuSTAR} to fit independently.  We attempted two more complex models (both with absorption fixed at the Galactic value as above). First, we considered a broken power law, with steeper low-energy and harder high-energy indices. This is similar to the model considered by \cite{wierzcholska2016}. The low- and high-energy indices are, respectively, ${2.52} \pm 0.07$ and ${1.81} \pm 0.08$, the break energy is at $5.2_{-0.5}^{+0.7}$ keV, and $\chi^{2}$ is 351 for 328 PHA channels. We note that the break energy in our fit is somewhat lower than that determined by \cite{wierzcholska2016}, but this is likely due to a different choice of bandpass, size of the source extraction region, and precise location of the region of the detector from which the background was subtracted.\par
\begin{figure}[htbp!]
\centering
\includegraphics[width=8cm]{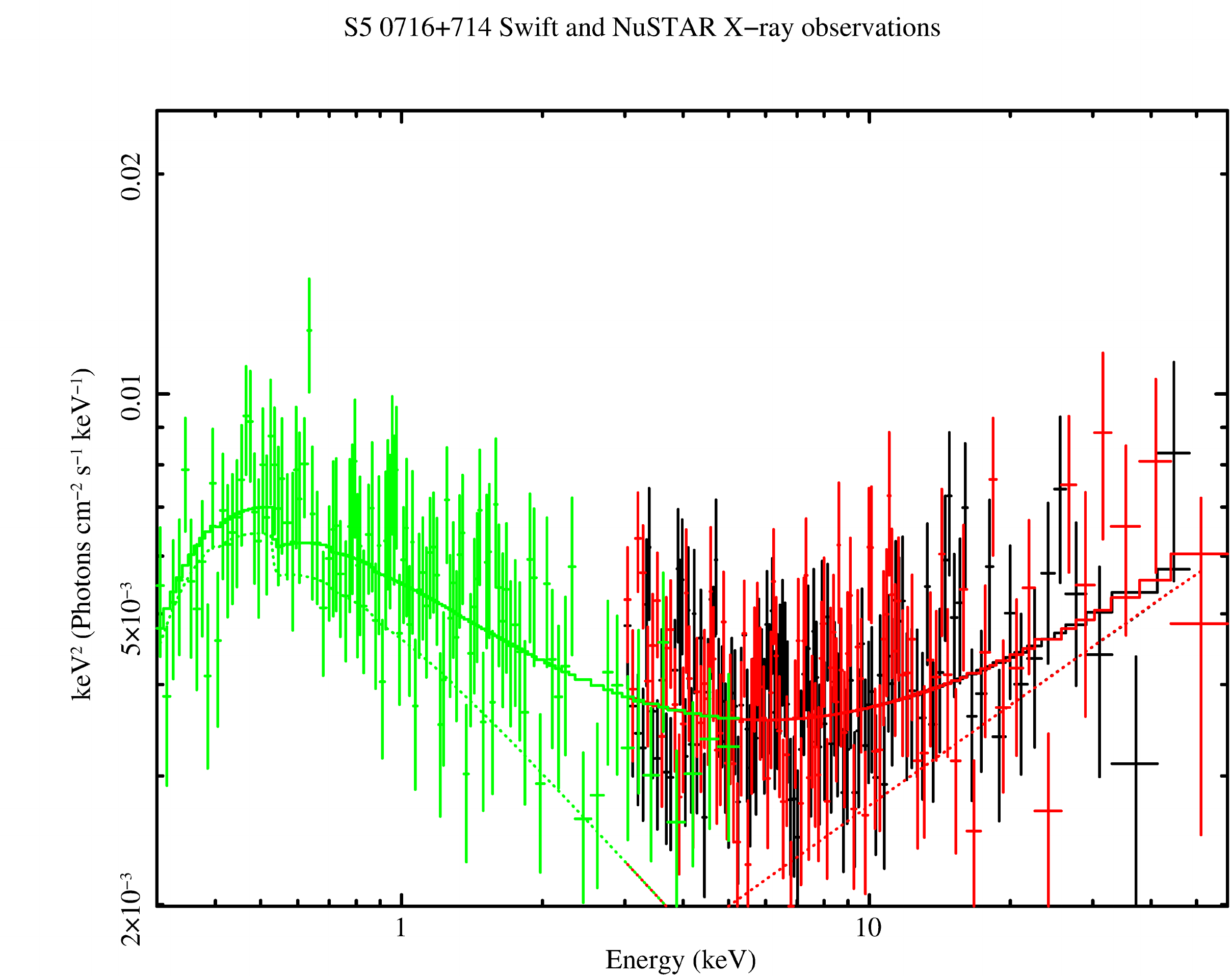}
\caption{Unfolded X-ray spectrum of S50716+714 derived from simultaneous fitting of the contemporaneous \textit{Swift}-XRT and \textit{NuSTAR} data obtained on 24 January 2015 (MJD~57046). The adopted model is a sum of two power laws. \textit{Swift}-XRT data are plotted in green, while the two \textit{NuSTAR} modules are plotted in red and black, respectively. The ``valley'' between the two main broad-band spectral peaks is in the X-ray band.}
\label{fig:nustar}
\end{figure}

We also attempted a double power-law representation of the data, also modified by Galactic absorption as above: here, the resulting spectrum is a sum of two power-law models, and is probably more physically motivated than a broken power law. The fit returns $\chi^{2}  = 352$ for 328 PHA bins with a low-energy index of ${2.62} \pm 0.16$ and a high-energy index of ${1.41 \pm 0.22}$. Since this model can represent a physically sensible superposition of two separate components, we express a preference for the two-power-law spectral form. With this model, the $2 - 10$ keV flux is $(9.7 \pm 0.7) \times 10^{-12}$ erg cm$^{-2}$ s$^{-1}$. We note here that the most reasonable interpretation of such a 2-component spectral shape is that we witness a contribution of two separate components, namely the ``tail'' of the low-energy component (presumably produced by the synchrotron process) and the onset of the high energy component (presumably due to the IC process). We plot the unfolded spectrum of the Swift XRT and \textit{NuSTAR} data observed on 24 January 2015 (MJD~57046) and fit to the two-power-law model in Fig.~\ref{fig:nustar}.
\section{VHE Differential Energy Spectrum and EBL deabsorbtion}
\begin{figure}[htbp!]
\centering
    \includegraphics[width=0.5\textwidth]{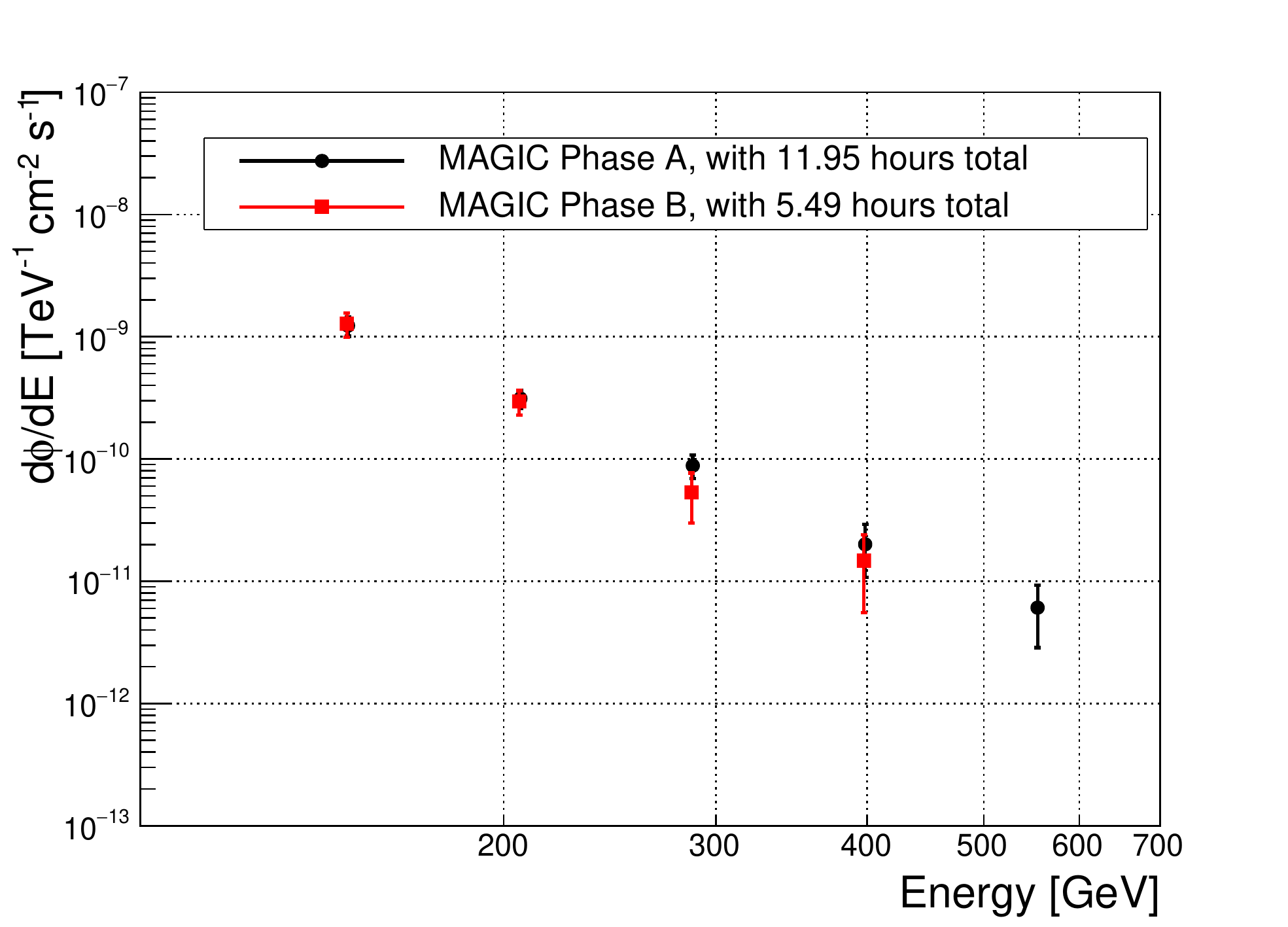}  
    \caption{Unfolded observed differential energetic spectra by MAGIC for Phase A (black full dots) and Phase B (red full squares). Parameters for the spectra (including the ones regarding the intrinsic EBL deabsorbed spectra with \cite{Dominguez2011} model) are reported in Table~\ref{table:VHEspectrum}. }
    \label{fig:Spectrum_MAGIC}
%\end{figure}
\end{figure}

The VHE $\gamma$-rays from distant blazars can interact with the optical-UV photons from the extragalactic background light (EBL)\citep{gould1967,stecker1992} via pair production, resulting in an attenuation of the intrinsic VHE spectrum. Finite resolution of the instrument will also modify the intrinsic spectrum. Unfolding techniques are adopted in the MARS code to unfold the observed spectrum from the instrument response. The differential spectrum of S5~0716+714 is shown in Fig.~\ref{fig:Spectrum_MAGIC} for a simple unfolding considering instrumental response only (hereafter observed spectrum). An unfolding including also de-absorption from EBL with the \cite{Dominguez2011} model (hereafter the intrinsic spectrum) was also performed, and parameters of the observed and intrinsic spectra are reported in Table~\ref{table:VHEspectrum}. The EBL imprint on the $\gamma$-ray spectra from distant blazars could be used to constrain the EBL density, under some assumptions on the intrinsic spectrum of the source \citep[see e.g.][]{ackermann2012,abramowski2013}. 
\begin{table*}[htbp!]
\caption{VHE spectrum parameters for a PWL fit}             
\label{table:VHEspectrum}      
\centering          
\begin{tabular}{ l c c c c c}     % 4 columns 
\hline\hline 
& $f_{0} (\mathrm{cm}^{-2} \ \mathrm{s}^{-1} \ \mathrm{TeV}^{-1})$ & $\Gamma$  & $\chi^{2}$/n.d.f. & $P$  & $E$ (GeV)\\  % table heading       
\hline
\hline
\noalign{\smallskip}
\multicolumn{4}{c}{-Phase A-} \\
\noalign{\smallskip}
\hline 
\hline   
observed    &  $(2.75 \pm 0.25_{stat} \pm 0.30_{sys}) \times 10^{-10}$  & $4.08 \pm 0.22_{stat} \pm 0.15_{sys}$ & $0.11/3$ & $0.99$  & 127.2-659.1 \\ % inserting body of the table
intrinsic & $(4.85 \pm 0.45_{stat} \pm 0.53_{sys}) \times 10^{-10}$  & $2.73 \pm 0.28_{stat} \pm 0.15_{sys}$  & $1.68/3$  & $0.64$ & 127.2-659.1\\                 
\hline
\hline
\noalign{\smallskip}
\multicolumn{4}{c}{- Phase B -} \\
\noalign{\smallskip}
\hline
\hline  
observed    & $(7.94 \pm 1.27_{stat} \pm 0.87_{sys}) \times 10^{-10}$  & $4.64 \pm 0.49_{stat} \pm 0.15_{sys}$ & $0.17/2$  & $0.92$ & 127.2-474.3\\ % inserting body of the table
intrinsic & $(1.06 \pm 0.17_{stat} \pm 0.11_{sys}) \times 10^{-10}$  & $3.65 \pm 0.53_{stat} \pm 0.15_{sys}$ & $0.23/2$  & $0.89$ & 127.2-474.3\\
\hline                
\end{tabular}
\end{table*}
The differential VHE spectra, observed as well as EBL-corrected using the model of \cite{Dominguez2011}, can be described by a power-law : 
\begin{center}
  \begin{equation}
  \centering
    \frac{dF}{dE}=f_0 \left(\frac{E}{150\,\mathrm{GeV}}\right)^{-\Gamma},
    \label{observed_spectrum}
  \end{equation}
\end{center}
where the normalization constant $f_0$, the spectral index $\Gamma$ , the goodness of the fit ($\chi^2$/n.d.f. and probability $P$), the energy range of the fit $E$ are indicated in Table~\ref{table:VHEspectrum} for Phase A and Phase B data respectively.
\subsection{Redshift estimation}
The simultaneous spectra from MAGIC and {\it Fermi}-LAT were used to estimate the redshift of the source. We apply the method presented in \cite{prandini2010, prandini2011} based on the assumption that the slope of the VHE spectrum corrected for EBL absorption should not be harder than the one measured by {\it Fermi}-LAT at lower energies. The redshift at which the two slopes match, z*, is then used as upper limit estimate of the source distance if there is no spectroscopic redshift available. If we apply the method to the data presented here assuming the \cite{Franceschini2008} EBL model, a 2 $\sigma$ upper limit on the redshift of 0.598 is found.\par
The empirical formula proposed in \cite{prandini2011} applied to this data gives as most probable value for the redshift $z_{rec} = 0.31 \pm 0.02_{stat} \pm 0.05_{sys}$, where the first error is related to the statistical errors of {\it Fermi}-LAT and MAGIC slopes, while the second error is the error of the method itself, as estimated in \cite{prandini2011}. This value is in agreement with the ones given in literature by \cite{nilsson2008,danforth2013}. The value of $z = 0.31 \pm 0.08$ found in \cite{nilsson2008} was based on the photometric detection of the host galaxy, while the $z < 0.322$ (95\% confidence) result reported by \cite{danforth2013} was obtained by detection of Ly-$\alpha$ systems in the ultra-violet spectrum of the source. For the SED modelling (see next section), we used the redshift value of 0.26 as in \cite{anderhub2009}. This value is within the errorbars of the redshift determined here as well as within other observations (see the Introduction).
\section{Broadband Spectral Energy Distribution}
\begin{figure*}[htbp!]
\centering
         \begin{subfigure}[t]{8.5cm}
               \includegraphics[width=8.5 cm]{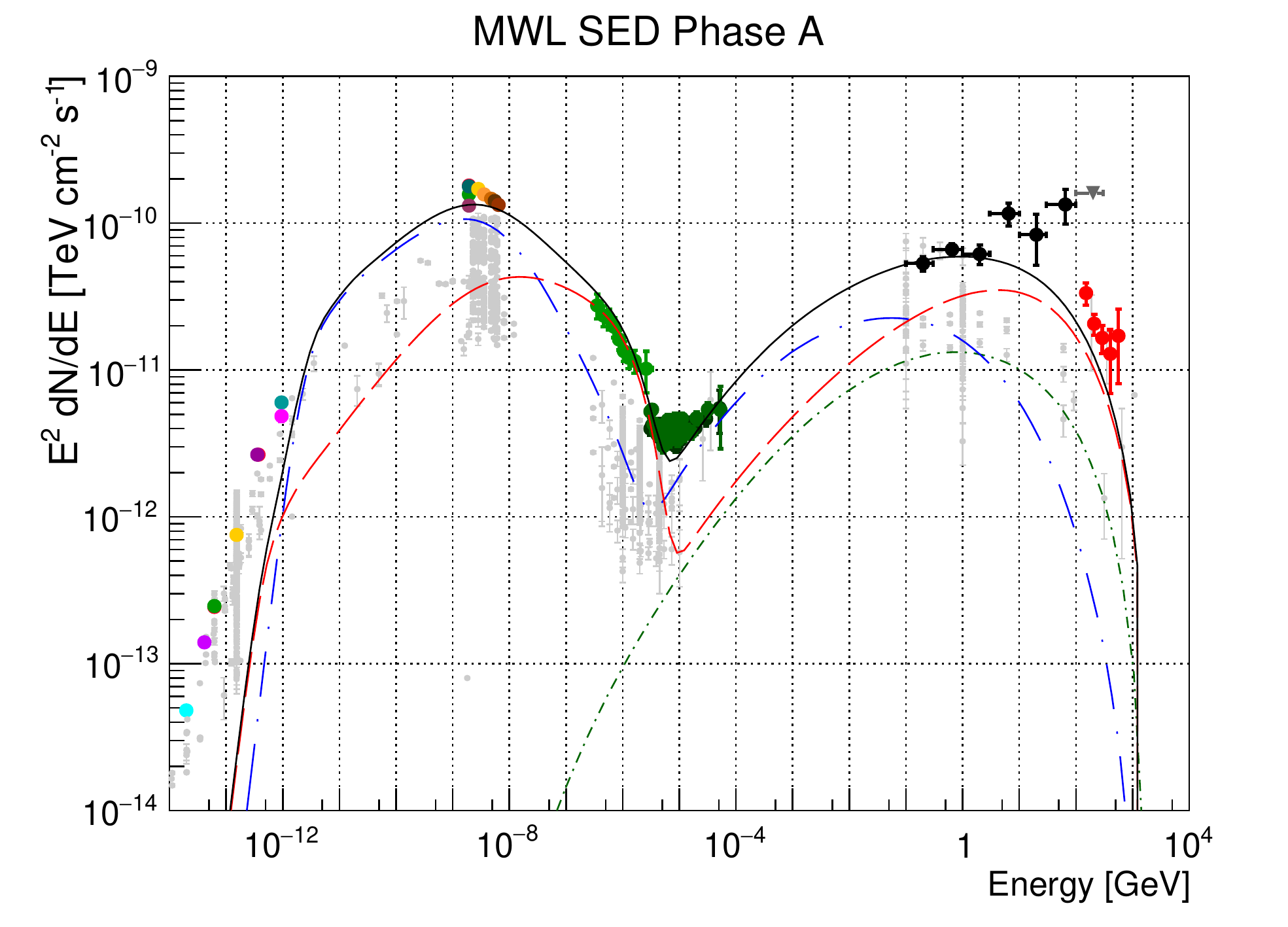}
               \label{fig:SED_Jan}
          \end{subfigure}
          \quad
          \begin{subfigure}[t]{8.5cm}
                \includegraphics[width=8.5 cm]{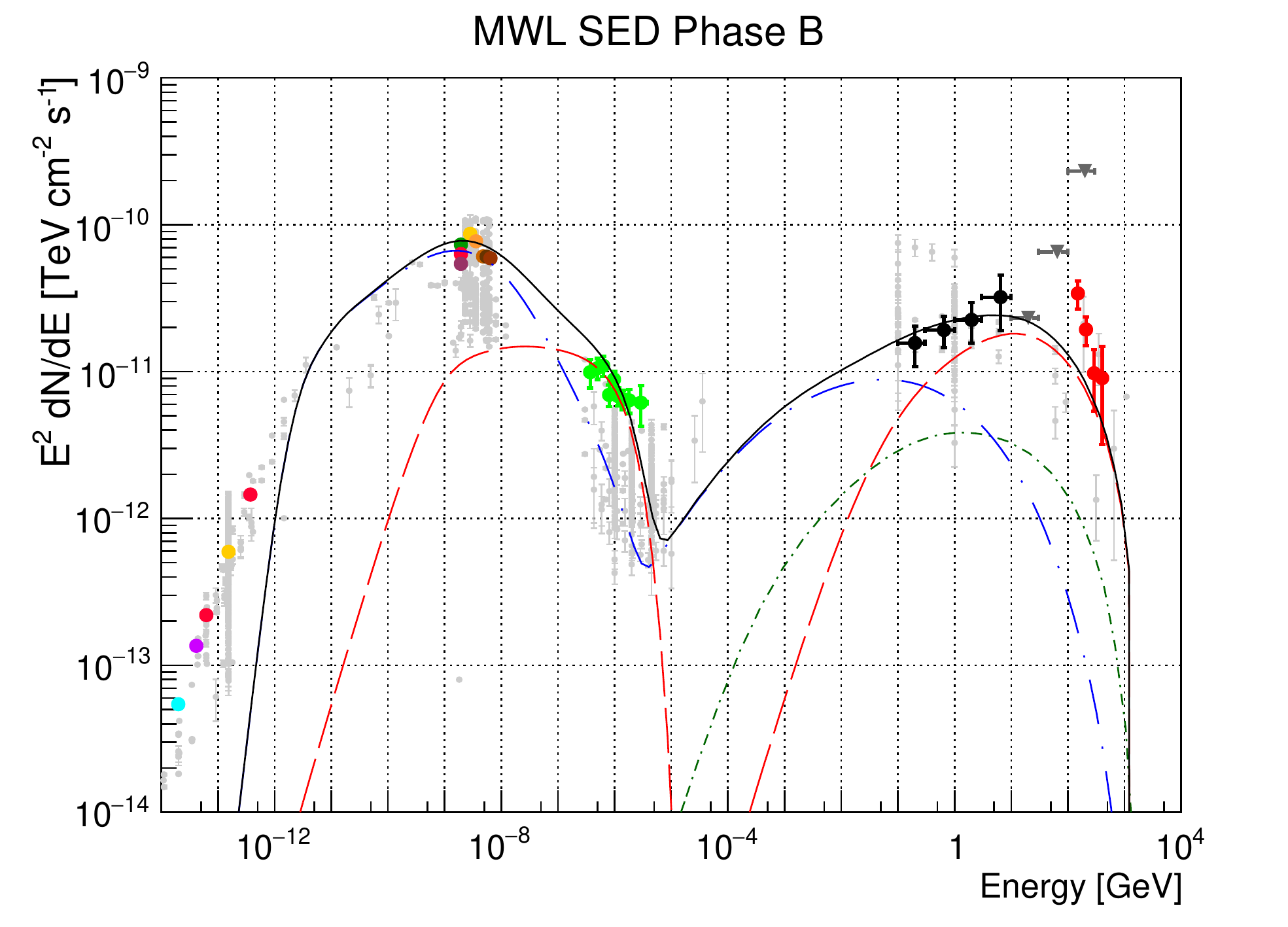}
                \label{fig:SED_Feb}
           \end{subfigure}     
 \caption{MWL Spectral Energy distributions for Phase A and Phase B. Archival data form ASDC are shown in grey. The two components (blobs representing a moving emission feature and a recollimation shock, see text) are shown with blue and red dashed lines. The green line is the emission that is a result of interaction between these two blobs and the black solid line the sum of these three components. The red full circles represent the intrinsic (EBL deabsorbed according to \citep{Dominguez2011}) MAGIC SED used in the model. For data taken in the radio and optical band the error bars are smaller than the size of the marker.}
 \label{fig:MWL_SED} 
\end{figure*}
The multi-wavelength SEDs for Phase A and B respectively are presented in Fig.~\ref{fig:MWL_SED}. Archival data from ASDC (ASI Science Data Center)\footnote{http://www.asdc.asi.it/} are shown in grey. When the source was detected for the first time in the VHE range \citep{anderhub2009} by MAGIC, the only available simultaneous multi-wavelength data were coming from KVA (optical) and \textit{Swift} (X-rays) and there were no constraints on the second bump beyond the MAGIC data. Nevertheless the very soft X-ray spectra belonging to the synchrotron component, combined with the high VHE $\gamma$-ray flux challenged the simple one-zone SSC model as it would require a very high flux of $\gamma$-rays around 10 GeV, higher than has been observed from the source with \textit{Fermi}-LAT or its precursor EGRET. This condition for one-zone persists also for the new data, but now we actually have simultaneous data in this energy range from \textit{Fermi}-LAT and we find that we cannot describe the observed broadband SED with the one-zone SSC model during this flaring period.\par
While a one-zone SSC model can match the observed \textit{Swift}-XRT+\textit{NuSTAR} spectrum as a transition between synchrotron and IC components, and simultaneously the $\gamma$-ray data from \textit{Fermi}-LAT and MAGIC, it tends to under-reproduce the observed optical flux. This has been independently verified using the BLAZAR code \citep{moderski2003}.\par
\begin{table*}[htbp!]
\caption{Input parameters for the emission models of S5~0716+714}             
\label{table:ParModeling}      
\centering          
\begin{tabular}{ l c c c c c c c c c c}     % 10 columns 
\hline\hline 
&$\gamma_{min}$ & $\gamma_{b}$  & $\gamma_{max}$&  $n_{1}$ & $n_{2}$ & $B$ (G) & $K$ & $R$ (cm) & $\delta$ & $z$  \\  % table heading       
\hline
\hline
\noalign{\smallskip}
\multicolumn{10}{c}{-Phase A-} \\
\noalign{\smallskip}
\hline 
\hline                    
"blob1"  & $100$  & $1.3\times10^{4}$ & $2\times10^{5}$  & $1.95$ & $3.5$  & $0.1$    & $3.5\times10^{3}$ & $2.6\times10^{16}$ & $25$ & $0.26$\\ % inserting body of the table
"blob2" & $300$  & $1\times10^{4}$ & $1.5\times10^{5}$  & $2.32$ & $4.6$  & $0.12$    & $1.6\times10^{4}$ & $6.5\times10^{16}$ & $25$ & $0.26$\\\\

\hline
\hline
\noalign{\smallskip}
\multicolumn{10}{c}{- Phase B -} \\
\noalign{\smallskip}
\hline
\hline  
"blob1"  & $4\times10^{3}$   & $8\times10^{3}$ & $2\times10^{5}$  & $1.9$ & $3.2$  & $0.09$  & $9\times10^{3}$ & $1.35\times10^{16}$ & $25$ & $0.26$\\ % inserting body of the table
"blob2" & $300$  & $1\times10^{4}$ & $1.5\times10^{5}$  & $2.32$ & $4.6$  & $0.12$  & $1\times10^{4}$ & $6.5\times10^{16}$  & $25$ & $0.26$\\

\hline                
\end{tabular}
\end{table*}
Based on the multiwavelength data in Section 3 and VLBA data in Section 4, we use two-zone model to describe the SED in Phase A and Phase B. We use two blobs close to each other to represent a situation where a superluminal knot (blob 1) is interacting a recollimation shock region (blob 2). The Phase A SED represents a snapshot of a time when the knot enters the recollimation shock region and the Phase B SED a time when the knot has exited the recollimation region. We model the two blobs with the framework similar to one presented for flat-spectrum radio quasar PKS~1222+216 in \cite{tavecchio2011}, modified for the case of no external seed photons as in \cite{Aleksic2014} for PKS~1424+240. Unlike the case of PKS~1424+240, in our case the two emission regions are very close to each other and they provide seed photons for inverse Compton scattering for each other. The model blobs have broken power-law electron spectra ($\gamma_{min}$, $\gamma_{b}$ and $\gamma_{max}$ are the minimum, break and maximum Lorentz factors respectively; $n_{1}$ and $n_{2}$ are the low and high energy slope of the smoothed power law electron energy distribution), magnetic field $B$, normalization of the electron distribution $K$, radius of the emission region $R$ and Doppler factor $\delta$.\par
We also use the observed variability behaviour as a guide on how the parameters change between Phase A and Phase B. As discussed in Section 3, Phase A and Phase B have different variability behaviours; while Phase A consists of a flare in all bands, in Phase B the activity is constrained to the X-ray and VHE $\gamma$-ray bands. To limit the number of free parameters, we fix the larger component (which is representing the recollimation shock) to have most of the parameters the same in Phases A and B. We only change $K$ to a lower value to represent the general lower state in optical and GeV $\gamma$-rays of Phase B. We then find parameters for the smaller emission region to describe the observed SEDs in Phase A and B separately. In both panels of Fig.~\ref{fig:MWL_SED} the "blob1" component is represented by the red dashed line. The "blob2" component of the model is represented by the blue dash-dotted line in both panels, and the emission resulting from the interaction of two components is reported with a green line.\par
We report a set of parameters we found to give a reasonable description (but see below) of the observed SED in Table~\ref{table:ParModeling} for the both phases. The set of parameters we present is not unique, but the parameters used are within the range typically found for TeV blazars and also for the ones found for PKS~1424+240 using the similar modelling setup.\par
Even if this simple two-zone model provides a better representation of the observed data from radio to VHE $\gamma$-rays respect to a one-zone model, the model line is slightly lower than the $\gamma$-ray fluxes in the range of 10-100~GeV even though within the systematic uncertainties of the data. The data in this energy range suggests a rather sharp feature, which is impossible to reproduce with this simplistic model we used. In general sharp features require presence of external seed photons such as used e.g. by \cite{bottcher2013} to model the SED of the source in a lower state. However, there is no observational evidence for such an external seed photon field from optical spectroscopy nor from the scenario we presented for the flaring behaviour within this epoch therefore no such a component was added to the modelling. \par 
There are also other possible two-zone model setups, such as a spine-sheath model \citep{ghisellini2005}, where a slower sheath of the jet surrounds a faster spine. For the previous VHE $\gamma$-ray flaring epoch a spine-sheath mode \citep{ghisellini2005}, was shown to provide a reasonable fit to the SED data \citep{anderhub2009,tavecchio2009}. We tested the spine-sheath model for the spectral energy distributions shown here and found acceptable agreement with the SED data. This emphasizes that SED data alone are not enough to separate different two-zone models, but must be combined with constraints from VLBA and light curve variability.\par
According to the scenario described above, which is supported by the dedicated VLBI study we performed in Sec.~4, we explain the extremely fast rotation of $\sim360^{\circ}$ as produced by turbulence in the interaction between a superluminal knot and a stationary feature near the core. Being dependent on the orientation of the shock and the magnetic field threading it, EVPA provides a unique tool to understand the acceleration mechanisms and behavior of the shocked plasma. Recent studies on EVPA swings larger than $180^{\circ}$ simultaneous with HE $\gamma$-ray emission \citep{marscher2008,marscher2010,abdo2010} have been interpreted as additional evidence for a helical magnetic field structure. The existing models focusing on the description of the synchrotron polarization features \citep[e.g.][]{lyutikov2005} apply a simple and time-independent power-law electron spectrum not taking into account possible predictions for the resulting HE emission. Our model, on the other hand, does not include a detailed geometry of the magnetic field and the angle-dependent synchrotron emissivity and polarization. At the moment only two models may represent the SED of blazars together with their synchrotron polarization features, including rotations of the EVPA: the HMFM (Helical Magnetic Field Model) \citep{zhang2014} and the TEMZ (Turbulent, Extreme Multi-Zone) \citep{marscher2014}. In the HMFM large polarization angle rotations by $\gtrsim 180^{\circ}$ are explained with the passage of a moving shock through a region with a highly disordered field: the compression of the shock orders the field partially, but this ordering is seen at different depths as time advances owing to light-travel delays. This leads to an apparent rotation of the polarization of $180^{\circ}$ per shock. In the TEMZ model randomness in the magnetic field direction in different turbulent cells can cause observed rotations in the linear polarization vector, even fast as the one observed in our case. Turbulence in general gives at different times "clusters" with small EVPA variation, relatively smooth EVPA rotations, step-wise EVPA changes, and random fluctuations. The behavior of S5~0716+714 in Phase A is consistent with this scenario.

\section{Summary and Conclusions}
The BL~Lac object S5~0716+714 has been studied in a multi-wavelength frame from radio to the VHE $\gamma$-ray band. In January 2015 an unprecedented outburst of S5~0716+714 was registered in all energy bands, from low frequency radio to VHE, and after almost a month another high state was detected by the MAGIC and \textit{Swift}-XRT instruments only. We divide the data into two phases (Phase A from 18 to 27 January 2015 -- MJD~57040 to MJD~57050, and Phase B from 12 to 17 February 2015 --MJD~57065 to MJD~57070) that represent very different characteristics, allowing a deep study of the broadband SEDs. \par
The broadband flaring activity period of Phase A coincides with the passage of a moving feature through a stationary feature (at $\sim$0.1 mas). We have found a very fast change in the electric vector position angle during Phase A. The $>~$400 degree swing in the optical EVPA is explained here as the passage of a superluminal knot through a stationary feature near the radio core. The VHE emission is then found originating in the entrance and exit of a superluminal knot in and out a recollimation shock in the inner jet. This suggests that shock-shock interaction in the jet seems to be responsible for the observed flares and EVPA swing. \par
The jet behaviour, studied with VLBA-BU-BLAZAR data, is in agreement with the scenario described in \cite{rani2015}, suggesting a connection between jet kinematics and the observed broadband flaring activity. More precisely, the $\gamma$-ray emission in the HE and VHE bands is attributed to a shock in the helical jet downstream of the core, closely followed by an optical and X-ray outburst in the core. The presence of low radio activity, observed during Phase A, was not reported in April 2008, when MAGIC observed the source in the VHE range for the first time \citep{anderhub2009}, but it could be a delayed response of a previous less intense flare, as observed in the past in the same source, when between optical/$\gamma$ flares lagged the radio counterparts almost two months \citep{rani2013a,rani2014}. \par
The first peak in the VHE $\gamma$-ray emission takes place $\sim$~2 days after the very fast EVPA rotation and the second $\sim$~18 days after the new knot has been emerged from the VLBA core. This is a strong indication that the VHE $\gamma$-ray emission is associated to a component entering and exiting the core region.\par
The broadband SEDs, for the first time including MAGIC and \textit{Fermi}-LAT simultaneous data and the quasi-simultaneous \textit{NuStar} data, could not be described by a simple one-zone model. Instead we used a two-zone model, where two spherical blobs are co-spatial and provide seed photons to each other. This modelling setup provides an acceptable description of the spectral energy distributions in Phase A and B, even if it is certainly an over-simplified presentation of the true physical processes taking in place when superluminal knots enter and exit the recollimation shock region.\par
Finally we also investigated the redshift of S5~0716+714. Using simultaneous data from MAGIC and {\it Fermi}-LAT the redshift was calculated to be $z = 0.31 \pm 0.02_{stat} \pm 0.05_{sys}$, confirming the value present in literature  based on the photometric detection of the host galaxy \citep{nilsson2008} and the more recent upper limit from a direct detection \citep{danforth2013}.\par
S50716+714 is an intermediate BL~Lac object, and only a handful of these sources have been detected in VHE $\gamma$-rays. In almost all detections of VHE $\gamma$-rays, activity in other bands (optical and/or HE $\gamma$-rays) has been seen, but our very comprehensive dataset provided a unique insight on how these VHE $\gamma$-ray flares are connected to the activity in the jet.\par
In the end of December 2017 S5~0716+714 was flaring again in VHE $\gamma$-rays \citep{atel11100}. It will be interesting to see if the recognized patterns will repeat also during this ongoing flaring period. This will be studied in a future paper.

%______________________________________________________________

%\section{Conclusions}

 %  \begin{enumerate}
  %    \item 
   %\end{enumerate}

\begin{acknowledgements}
%MAGIC
We would like to thank the Instituto de Astrof\'{\i}sica de Canarias for the excellent working conditions at the Observatorio del Roque de los Muchachos in La Palma. The financial support of the German BMBF and MPG, the Italian INFN and INAF, the Swiss National Fund SNF, the ERDF under the Spanish MINECO (FPA2015-69818-P, FPA2012-36668, FPA2015-68378-P, FPA2015-69210-C6-2-R, FPA2015-69210-C6-4-R, FPA2015-69210-C6-6-R, AYA2015-71042-P, AYA2016-76012-C3-1-P, ESP2015-71662-C2-2-P, CSD2009-00064), and the Japanese JSPS and MEXT is gratefully acknowledged. This work was also supported by the Spanish Centro de Excelencia ``Severo Ochoa'' SEV-2012-0234 and SEV-2015-0548, and Unidad de Excelencia ``Mar\'{\i}a de Maeztu'' MDM-2014-0369, by the Croatian Science Foundation (HrZZ) Project IP-2016-06-9782 and the University of Rijeka Project 13.12.1.3.02, by the DFG Collaborative Research Centers SFB823/C4 and SFB876/C3, the Polish National Research Centre grant UMO-2016/22/M/ST9/00382 and by the Brazilian MCTIC, CNPq and FAPERJ.
%Fermi
The \textit{Fermi} LAT Collaboration acknowledges generous ongoing support from a number of agencies and institutes that have supported both the development and the operation of the LAT as well as scientific data analysis. These include the National Aeronautics and Space Administration and the Department of Energy in the United States, the Commissariat \`a l'Energie Atomique
and the Centre National de la Recherche Scientifique / Institut National de Physique Nucl\'eaire et de Physique des Particules in France, the Agenzia Spaziale Italiana and the Istituto Nazionale di Fisica Nucleare in Italy, the Ministry of Education, Culture, Sports, Science and Technology (MEXT), High Energy Accelerator Research Organization (KEK) and Japan Aerospace Exploration Agency (JAXA) in Japan, and the K.~A.~Wallenberg Foundation, the Swedish Research Council and the Swedish National Space Board in Sweden. Additional support for science analysis during the operations phase is gratefully acknowledged from the Istituto Nazionale di Astrofisica in Italy and the Centre National d'\'Etudes Spatiales in France. This research was supported by an appointment to the NASA Postdoctoral Program at the Goddard Space Flight Center, administered by Universities Space Research Association through a contract with NASA.%end Fermi 
We thank the \textit{Swift} team duty scientists and science planners. The Mets\"ahovi team acknowledges the support from the Academy of Finland to our observing projects (numbers 212656, 210338, 121148, and others). The VLBA is an instrument of the Long Baseline Observatory. The Long Baseline Observatory is a facility of the National Science Foundation operated by Associated Universities, Inc. The Submillimeter Array is a joint project between the Smithsonian Astrophysical Observatory and the Academia Sinica Institute of Astronomy and Astrophysics and is funded by the Smithsonian Institution and the Academia Sinica. The OVRO 40-m monitoring program is supported in part by NASA grants NNX08AW31G, NNX11A043G and NNX14AQ89G, and NSF grants AST-0808050 and AST-1109911. The St. Petersburg University team acknowledges support from Russian Science Foundation grant 17-12-01029. The BU group acknowledges support by NASA under {\it Fermi} Guest Investigator grant NNX14AQ58G and by NSF under grant AST-1615796. Part of this work was done with funding by the UK Space Agency. The VLBA is an instrument of the National Radio Astronomy Observatory. The National Radio Astronomy Observatory is a facility of the National Science Foundation operated under cooperative agreement by Associated Universities, Inc. The PRISM (Perkins Re-Imaging SysteM) camera at Lowell Observatory was developed by K.\ Janes et al. at BU and Lowell Observatory, with funding from the NSF, BU, and Lowell Observatory. This paper makes use of data obtained with the 100\,m Effelsberg radio-telescope, which is operated by the Max-Planck-Institut f\"ur Radioastronomy (MPIfR) in Bonn (Germany). Part of this work is based on archival data, software or online services provided by the ASI Science Data Center (ASDC). PYRAF is a product of the Space Telescope Science Institute, which is operated by AURA for NASA. This paper is partly based on observations carried out with the IRAM 30m. IRAM is supported by INSU/CNRS (France), MPG (Germany) and IGN (Spain). IA acknowledges support by a Ramon y Cajal grant of the Ministerio de Economia, Industria y Competitividad (MINECO) of Spain. The research at the IAA-CSIC was supported in part by the MINECO through grants AYA2016-80889-P, AYA2013-40825-P, and AYA2010-14844, and by the regional government of Andalucia through grant P09-FQM-4784. The Liverpool Telescope is operated by JMU with financial support from the UK-STFC.

\end{acknowledgements}


\newpage

\begin{appendix} %First appendix

\section{Knot parameters for S5~0716+714 (July 2014-August 2015)}
\label{sec:appendix}
In Table~\ref{table:ParKnots} the parameters of components from the Section~\ref{sec:VLBI} are listed. They indicate the flux density, $S$, the relative RA and DEC with respect to the core, the distance from the core $r$, the position angle with respect to the core, $\Theta$, and the size of the component, $a$ (FWHM of the Gaussian) for every knot. Some knots present in Table~\ref{table:ParKnots} do not have any assigned name because they have been seen at 1-2 epochs only.

\begin{table*}[htbp!]
\caption{Knot parameters for S5~0716+714 (July 2014-August 2015)}             
\label{table:ParKnots}      
\centering          
\begin{tabular}{c c c c c c c c c}     % 9 columns 
\hline\hline 
Epoch &  MJD &  $S$(Jy)&  x(mas)& y(mas)& $r$(mas)& $\Theta$(deg)& $a$(mas)& Knot  \\  % table heading
\hline
\hline
\noalign{\smallskip}
\multicolumn{9}{c}{- 2014-Jul-28 -} \\
\noalign{\smallskip}
\hline 
\hline                    
2014.5726 & 56867 & 1.619 &  0.000 & 0.000 & 0.000  &  0.0 & 0.020 & A0  \\
2014.5726 & 56867 & 0.548 &  0.092 & 0.005 & 0.093  & 86.8 & 0.031 & A1  \\
2014.5726 & 56867 & 0.017 &  0.186 & 0.551 & 0.582  & 18.7 & 0.186 &     \\
\hline
\hline
\noalign{\smallskip}
\multicolumn{9}{c}{- 2014-Sep-23 -} \\
\noalign{\smallskip}
\hline 
\hline                    
2014.7288 & 56924 & 1.322 &  0.000 & 0.000 & 0.000  &  0.0 & 0.020 & A0  \\
2014.7288 & 56924 & 0.205 &  0.095 & 0.021 & 0.097  & 77.6 & 0.031 & A1  \\
2014.7288 & 56924 & 0.033 &  0.158 & 0.457 & 0.484  & 19.1 & 0.189 &     \\
\hline
\hline
\noalign{\smallskip}
\multicolumn{9}{c}{- 2014-Nov-15 -} \\
\noalign{\smallskip}
\hline 
\hline                    
2014.8740 & 56977 & 2.676 &  0.000 & 0.000 & 0.000  &  0.0  & 0.022  & A0     \\
2014.8748 & 56977 & 0.105 &  0.095 & 0.021 & 0.097  &  77.6 & 0.031  & A1     \\
2014.8740 & 56977 & 0.306 &  0.178 & 0.095 & 0.202  &  62.0 & 0.112  &  K14a  \\
\hline
\hline
\noalign{\smallskip}
\multicolumn{9}{c}{- 2914-Dec-5 -} \\
\noalign{\smallskip}
\hline 
\hline                    
2014.9288 & 56997 & 1.717  & 0.000 & 0.000 & 0.000  &  0.0 & 0.019 & A0\\ % inserting body of the table
2014.9288 & 56997 & 0.571  & 0.078 & 0.055 & 0.095  & 54.8 & 0.038 & A1\\
2014.9288 & 56997 & 0.183  & 0.206 & 0.139 & 0.248  & 56.1 & 0.094 & K14a\\
2014.9288 & 56997 & 0.019  & 0.163 & 0.440 & 0.469  & 20.3 & 0.274 &       \\
\hline
\hline
\noalign{\smallskip}
\multicolumn{9}{c}{- 2014-Dec-29 -} \\
\noalign{\smallskip}
\hline
\hline  
2014.9945 & 57021 & 1.171  & 0.000 & 0.000 & 0.000 &   0.0 & 0.017 &  A0    \\
2014.9945 & 57021 & 0.168  & 0.093 & 0.077 & 0.121 &  50.5 & 0.023 &  A1    \\
2014.9945 & 57021 & 0.245  & 0.250 & 0.179 & 0.307 &  54.4 & 0.130 &  K14a  \\
2014.9945 & 57021 & 0.021  & 0.246 & 0.602 & 0.651 &  22.2 & 0.416 &        \\
\hline
\hline
\noalign{\smallskip}
\multicolumn{9}{c}{- 2015-Feb14 -} \\
\noalign{\smallskip}
\hline
\hline 
2015.1233 & 57067 & 1.901 & 0.000 & 0.000 & 0.000 &  0.0 & 0.013 & A0  \\
2015.1233 & 57067 & 0.314 & 0.093 & 0.080 & 0.123 & 49.2 & 0.065 & A1  \\
2015.1233 & 57067 & 0.122 & 0.232 & 0.222 & 0.321 & 46.3 & 0.155 & K14a\\
2015.1233 & 57067 & 0.036 & 0.236 & 0.527 & 0.577 & 24.1 & 0.412 &     \\
\hline
\hline
\noalign{\smallskip}
\multicolumn{9}{c}{- 2015-Apr-11 -} \\
\noalign{\smallskip}
\hline
\hline 
2015.2767 & 57123 & 2.094 & 0.000 & 0.000 & 0.000 &  0.0 & 0.017 & A0  \\
2015.2767 & 57123 & 0.373 & 0.069 & 0.056 & 0.089 & 51.3 & 0.057 & A1  \\
2015.2767 & 57123 & 0.045 & 0.163 & 0.169 & 0.235 & 44.1 & 0.100 & K14b\\
2015.2767 & 57123 & 0.089 & 0.287 & 0.390 & 0.484 & 36.3 & 0.179 & K14a\\
\hline
\hline
\noalign{\smallskip}
\multicolumn{9}{c}{- 2015-May-11 -} \\
\noalign{\smallskip}
\hline
\hline 
2015.3589 & 57153 & 1.581 & 0.000 & 0.000 & 0.000 &  0.0 & 0.023 & A0  \\
2015.3589 & 57153 & 0.309 & 0.074 & 0.065 & 0.098 & 48.7 & 0.063 & A1  \\
2015.3589 & 57153 & 0.032 & 0.204 & 0.165 & 0.262 & 50.9 & 0.095 & K14b\\
2015.3589 & 57153 & 0.069 & 0.286 & 0.407 & 0.498 & 35.1 & 0.201 & K14a\\
\hline
\hline
\noalign{\smallskip}
\multicolumn{9}{c}{- 2015-Jun-9 -} \\
\noalign{\smallskip}
\hline
\hline 
2015.4385 & 57182 & 0.947 & 0.000 & 0.000 & 0.000 &  0.0 & 0.018 & A0  \\
2015.4385 & 57182 & 0.308 & 0.106 & 0.098 & 0.144 & 47.5 & 0.067 & A1  \\
2015.4385 & 57182 & 0.039 & 0.229 & 0.191 & 0.298 & 50.3 & 0.069 & K14b\\
2015.4385 & 57182 & 0.054 & 0.320 & 0.437 & 0.542 & 36.2 & 0.189 & K14a\\
\hline
\hline
\noalign{\smallskip}
\multicolumn{9}{c}{- 2015-Jul-2 -} \\
\noalign{\smallskip}
\hline
\hline 
2015.5014 & 57205 & 1.612 & 0.000 & 0.000 & 0.000 &  0.0 & 0.021 & A0  \\
2015.5014 & 57205 & 0.124 & 0.020 & 0.060 & 0.063 & 18.7 & 0.000 & K15 \\
2015.5014 & 57205 & 0.144 & 0.075 & 0.096 & 0.122 & 37.9 & 0.084 & A1  \\
2015.5014 & 57205 & 0.041 & 0.253 & 0.204 & 0.325 & 51.1 & 0.084 & K14b\\
2015.5014 & 57205 & 0.057 & 0.322 & 0.384 & 0.531 & 40.0 & 0.141 & K14a\\
2015.5014 & 57205 & 0.016 & 0.376 & 0.665 & 0.764 & 29.5 & 0.401 &     \\
\hline
\hline
\noalign{\smallskip}
\multicolumn{9}{c}{- 2015-Aug-1 -} \\
\noalign{\smallskip}
\hline
\hline 
2015.5836 & 57235 & 1.865 & 0.000 & 0.000 & 0.000 &  0.0 & 0.019 & A0  \\
2015.5836 & 57235 & 0.238 & 0.045 & 0.090 & 0.101 & 26.3 & 0.033 & K15 \\
2015.5836 & 57235 & 0.245 & 0.087 & 0.126 & 0.153 & 34.6 & 0.083 & A1  \\
2015.5836 & 57235 & 0.032 & 0.241 & 0.278 & 0.368 & 41.0 & 0.107 & K14b\\     
2015.5836 & 57235 & 0.018 & 0.388 & 0.455 & 0.598 & 40.4 & 0.119 & K14a\\
2015.5836 & 57235 & 0.020 & 0.315 & 0.667 & 0.738 & 25.3 & 0.398 &     \\
\hline
\hline                 
\end{tabular}
\end{table*}

\end{appendix}

\end{document}